\documentclass[lettersize,journal]{IEEEtran}
\usepackage{amsmath,amsfonts}
\usepackage{algorithmic}
\usepackage{algorithm}
\usepackage{array}
\usepackage{subfig}
\usepackage{textcomp}
\usepackage{stfloats}
\usepackage{url}
\usepackage{verbatim}
\usepackage{graphicx}
\usepackage{cite}

\usepackage{imakeidx}
\makeindex

\usepackage{amscd,amsmath,amssymb,amsfonts,latexsym,mathrsfs,amsthm}
\usepackage{graphicx} 
\usepackage{underscore}
\usepackage{graphics,epsfig}
\usepackage{graphics,epsfig}
\usepackage[english]{babel}
\usepackage{amsmath}
\usepackage{amsfonts}
\usepackage{amssymb,amsthm}
\usepackage{bm}
\usepackage{mathrsfs}
\usepackage[usenames]{color}
\usepackage{rotating}
\usepackage{multicol}

\usepackage{hyperref}
\usepackage{url}

\theoremstyle{remark}

\usepackage{flushend}
\usepackage{verbatim}
\usepackage{tabularx}
\usepackage{multirow} 
\usepackage{arydshln}
\usepackage{ulem}

\UseRawInputEncoding

\begin{document}

\title{
An exhaustive variable selection study for linear models of soundscape emotions: rankings and Gibbs analysis }

\author{R. San Mill{\'a}n-Castillo$^*$, L. Martino$^*$, E. Morgado$^*$, F. Llorente$^\dagger$   \\
$^*$ Dep. of Signal Theory and Communications, Universidad Rey Juan Carlos (URJC), Madrid, Spain. \\
$^\dagger$ Dep. of  Statistics, Universidad Carlos III de Madrid (UC3M), Madrid, Spain.} 



\maketitle

\begin{abstract}
In the last decade, soundscapes have become one of the most active topics in Acoustics, providing a holistic approach to the acoustic environment, which involves human perception and context.  Soundscapes-elicited emotions are central and substantially subtle and unnoticed (compared to speech or music). Currently, soundscape emotion recognition is a very active topic in the literature.
We provide an exhaustive variable selection study (i.e., a selection of the soundscapes indicators)  to a well-known dataset (emo-soundscapes). We consider linear soundscape emotion models for two soundscapes descriptors: arousal and valence.
 Several ranking schemes and procedures for selecting the number of variables are applied. We have also performed an alternating optimization scheme for obtaining the best sequences keeping fixed a certain number of features. Furthermore, we have designed a novel technique based on Gibbs sampling, which provides a more complete and clear view of the relevance of each variable. Finally, we have also compared our results with the analysis obtained by the classical methods based on p-values. As a result of our study, we suggest two simple and parsimonious linear models of only $7$ and $16$ variables (within the 122 possible features) for the two outputs (arousal and valence), respectively. The suggested linear models provide very good and competitive performance, with $R^2>0.86$ and $R^2>0.63$ (values obtained after a  cross-validation procedure), respectively.       
\end{abstract}

\begin{IEEEkeywords}
Soundscape emotion, variable selection, ranking methods, best sequence search, MCMC algorithms, Gibbs sampling.
\end{IEEEkeywords}

\maketitle

\section{Introduction}


Environmental noise is one of the most critical risks for population health and well-being. The World Health Organization has recently remarked that noise affects at least 100 million people, only in the European Union \cite{Who2018}. Generally, sound level monitoring and control are the common tools for managing the acoustic environment and sound quality remains dismissed. However, noise abatement is often unavailable or unsuitable in certain scenarios like cities, or does not necessarily result in an approving appraisal of final soundscapes \cite{van2018evolution}. Hence, ``quiet areas" are a new perspective that focuses on the acoustic quality more than on the sound level, and which are being even regulated in the European Union \cite{european2014good}.
\newline
This vision is limited since it is not accountable for people's experiences in different acoustic environments. Soundscapes provide an alternative and holistic approach to assess human perception, acoustic environments, and context, beyond the concept of noise \cite{schafer1993soundscape}.
 Thus, this subjective evaluation depends on physical, psychological, social, and even cultural estimators and their complex interactions. 
\newline
\newline
In the last decade, soundscapes have become one of the most active topics in acoustics. In fact, the number of related research projects and scientific articles grows exponentially \cite{aletta2018current}. Research requires a sizeable sample of participants in surveys and a considerable amount of locations. These intensive and time-consuming resources may limit the soundscape approach \cite{lunden2016urban}. Soundscape modeling might predict people's perception of acoustic environments at lower expenses \cite{lionello2020systematic}. In urban planning and environmental acoustics, the procedure consists of (a)  soundscapes recording, (b) calculation of acoustic and psychoacoustic indicators of the signals, (c) collecting other context indicators (e.g. visual information \cite{axelsson2010principal}), 
and (d)
ranking of soundscapes audio signals employing emotional descriptors. Finally, the model can be developed. 
 \newline
Soundscape-elicited emotions are substantially different from those related to music or speech because they are more subtle and unnoticed. Thus, soundscape emotion recognition (SER) requires further research to support perception and context descriptors \cite{fiebig2020assessments, ma2015human}. Additionally to environmental acoustics and urban planning, there is an increasing research interest in SER for certain domains like sound design in films and digital games \cite{lopes2017modelling}, or sonification in the {\it Internet of Things (IoT)} \cite{abri2021comparative}. Soundscapes descriptors are identified with perceived emotions and SER becomes a relatively new sub-field of research in affective computing \cite{fan2017emo}. Rusell's circumplex model can be applied to soundscapes \cite{axelsson2010principal,russell1980circumplex,davies2014soundscape, fan2017emo} by scaling the perception of soundscapes. Russell\textquotesingle s affect representation can be modeled with two main factors: {\it arousal} represents the eventfulness of the acoustic environment, and {\it valence} is the pleasantness ratio. Currently, arousal and valence are accepted as the principal and sufficient affective descriptors in research \cite{davies2014soundscape, vastfjall2003affective}, but there are different proposals to enhance soundscapes emotions evaluation with additional or different descriptors \cite{aletta2016soundscape}, and even to include emotion appraisal in procedures of the soundscapes standards \cite{fiebig2020assessments}.
\newline
\newline
Soundscape modeling has been extensively and recently reviewed in \cite{lionello2020systematic}. Soundscapes indicators (i.e. features), soundscapes descriptors (i.e. outputs), and employed prediction models and their performances are presented. Researchers have been approaching SER from a variety of perspectives, and the results are roughly comparable. However, the published literature shows some trends. Firstly, a large dataset leads to stable and well-performing models. Indicators that include psychoacoustic and perceptual information contribute to improving model performance. Finally, {\it non-linear models} (NLMs) seem to result in (slightly) more accurate models than {\it linear models} (LMs). However, NLMs approaches remain complex and challenging for researchers since the model development and the hyperparameters tuning might become demanding. Hence, LMs are usually the preferred choice although they could be often outperformed by NLMs strategies. Some of the predictive LMs provide poor performance ($R^2 = 0.18$) \cite{herranz2016progress}, whereas other LMs achieve very good performance ($R^2 = 0.9$) \cite{aumond2017modeling}. On the other hand, reported NLMs use  sophisticated machine learning techniques such as support vector machines (SVM), artificial neural networks (ANN), or random forests (RF) to name a few. They outperform slightly LMs in terms of prediction, e.g., regarding scores ($R^2 = 0.91$) \cite{hong2019study}. Nevertheless, LMs still appear as prevalent in this field, while research with NLMs seems to be just promising, so far.
\newline
A wide range of descriptors is modeled by a large array of indicators in a variety of scenarios. Thus, a general framework for comparison seems not to be established. One of the reasons is the scarcity of SER datasets that are publicly available. Emo-soundscapes database (EMO) \cite{fan2017emo} sets up a free and available dataset for SER comparison from 2017, which is focused on arousal and valence. Thus, other researchers have been exploring EMO as a reference. Firstly, \cite{fan2017emo} presents a baseline for EMO based on two independent SVMs, in order to to model both arousal and valence, selecting 39 features by a variance threshold. In \cite{abri2020predicting}, a comparison of four LM and four NLM is explored and a dimension reduction is performed by a principal components analysis (PCA).\footnote{In order to avoid confusions, it is important to remark that the dimension reduction obtained by a PCA is different from a dimension reduction obtained by applying a variable selection scheme. The  dimension reduction by PCA is obtained by suitable linear combinations of variables. These linear combinations can be considered as ``new variables'' (and/or meta-features). A variable selection technique just selects some of existing variables  trying to removing useless redundancy (without creating new features).} Furthermore, the authors in \cite{fan2017emo} also show that a RF model outperforms the rest of the models with only 25 features. In \cite{abri2021comparative}, a fine-tuned RF model with 14 features overcomes the previous RF model, and convolution neural networks (CNNs).  Deep learning techniques have been also applied to SER through CNN and 23 simplified mel-frequency cepstral coefficients (MFCC) in \cite{ntalampiras2020emotional}, and the combination with SVM (Transfer learning) in \cite{fan2018soundscape}. Promising results use up to 54 features by heuristic methods despite the limited samples of EMO. Thus, EMO is the selected dataset for our study, because it is a suitable and relevant reference. 
\newline
 The goal of this work is to  design simple
and interpretable SER linear models. Additionally,
this study offers an exhaustive feature selection framework
that helps researchers adjust their model errors with the
features’ importance and their relationships. Namely, we
provide an exhaustive variable selection study for LM,
considering classical and novel methodologies. Hence, the required resources for SER modeling become less laborious and the research community can employ the designed models to improve the knowledge about SER. A wide range of applications such as urban planning, noise monitoring, and sonification production might improve their performance based on SER models. Moreover, variable selection may lead to less significant computing resources. This helps to bring SER models to devices with real-time responsivity, such as the IoT framework. First of all, 
we divide the methods into two main parts: ranking of variables   and  selection of the effective number of variables.  This approach yields several benefits: 
 (a) allows a better understanding of the different techniques, (b)  allows the combination of different ranking and number selection schemes, and  (c) produces a more complete view of the variable selection problem.  We consider five different ranking methods and also compare the results to the classical ranking method based on p-values \cite{efroymson1960multiple,hocking1976biometrics}. 
 Moreover, we apply the best sequence search (keeping fixed the number of variables). For this purpose, we perform an alternating optimization method that allows us finding easily (at least) local modes. We repeat the procedure for several different runs for obtaining the global mode.
 \newline
  Last but not least, we design a pseudo-target density and a Gibbs sampling scheme which allows us having a complete view of the importance of the variables in terms of prediction error. The results of the Gibbs analysis support and clarify the results obtained previously by the ranking methods and the best sequence search. Some other considerations are only remarked by the Gibbs analysis. This novel technique can be applied for general variable selection purposes (not just for the specific database analyzed).  The overall study allows us to propose (a) {\it parsimonious}, (b) {\it interpretable}, and (c) {\it robust} linear models. Namely, we can focus on a few very relevant variables that are highlighted in all the performed analyses. We believe that these variables keep their relevance also in different databases (as also suggested by the cross-validation results). Moreover, focusing on a few variables helps the   interpretability of the resulting model.  
 \newline 
  In summary, we aim at developing well-performing, simple, robust and interpretable SER linear models. In order to achieve this objective, the contributions of the work are the following:
\begin{itemize}
\item We apply five different ranking methods to analyze the relevance of the variables in terms of prediction error in arousal and valence.
\item Additionally, we apply two more sophisticated methods to find the most relevant variables: 
 1) the best sequence search; and 2) a Gibbs sampling approach building a suitable target density based on prediction error (a technique developed for the first time in this work but which can be applied for general variable selection purposes).
 \item Regarding the selection of the effective number of variables, we apply several information criteria (such as the well-known AIC and BIC \cite{llorente2020review}) and a classical $p-$value based approach. The obtained results are also compared with the Gibbs analysis.
\item Based on the complete and exhaustive analysis of the previous methods, we offer two different linear models to predict arousal and valence from a very reduced number of variables and with low prediction error.
\end{itemize}

The rest of the work is organized as follows. Section \ref{sec_dataset} describes the dataset which is the object of our study. Section \ref{sec_MLR} presents  some background material describing the LM. Section \ref{sec_varanalysis} describes the different techniques that we will use for our analysis: ranking methods, the algorithm for best sequence search, and the Gibbs sampling analysis.  Then, Sections \ref{sec_results1} and \ref{sec_results2} show the results applied to the EMO database for the first output (arousal) and the second output (valence), respectively. Finally, in Section \ref{sec_conc}, we discuss some conclusions. The detailed results for both outputs are given in the \texttt{Supplementary material}.

\section{Database and model descriptions}

\subsection{Dataset Description} \label{sec_dataset}
This research explores the EMO that might be considered as the largest publicly available soundscape database with annotations of emotion labels, and the most bench-marked up to now \cite{fan2017emo}. EMO contains 1213 audio clips which are released under Creative Commons license from the Freesound collaborative audio platform \cite{fonseca2017freesound}. The Schafer\textquotesingle s taxonomy classifies the selected clips into six categories due to their generality and simplicity . The Schafer\textquotesingle s categories consider both the identification of the source and the listening context \cite{schafer1993soundscape, fan2017emo} : natural sounds (e.g. birds, wind), human sounds (e.g. laugh, shouts), sounds and society (e.g. party, store), mechanical sounds (e.g. engine, factory), quiet and silence (e.g. quiet park, silent forest), and sounds as indicators (e.g. clock, church bells). EMO consists of 100 audio clips per category within a first subset (i.e. 600 audio clips) and 613 manually mixed sound of two or three categories of the first subset. A crowd-sourcing procedure provided data annotations of perceived arousal and valence, by a ranking-based questionnaire of a two clips pairwise comparison. Eventually, 1182 trusted annotators performed the required tasks with reasonable inter-subject reliability. 
\newline
Audio clips are monophonic and the sample rate was 44100 Hz, , which is widely considered a high quality standard for audio files. Monophonic recordings are sufficient to evaluate the eventfulness (arousal) and pleasantness (valence) of acoustic environments, among many other soundscapes indicators, according to \cite{xu2019soundscape}. EMO has employed both YAAFE \cite{mathieu2010yaafe} and MIRToolbox \cite{lartillot2008matlab} for the extraction of 122 normalized audio features, applying a 23 ms Hanning window with 50\% overlapping. There are three main groups of features of the audio signals:
\begin{itemize}
     \item {\bf Psychoacoustic features:} They are indexed 4, from 24 to 49, and 113, 114, 117, 118, 119.  These features represent perceptual (i.e., subjective) attributes of sounds such as level (i.e., loudness for overall level and MFCC for band-limited levels), spectrum (i.e., sharpness for high frequencies), and temporal and spectrum modulation (i.e., fluctuation).
    \item {\bf Time-domain features:} They are indexed from 1 to 7, and 22, 23, 52, 115, 116. These features represent the signal dynamics, such as classical estimators based on samples of the audio signal (i.e., energy, entropy of energy, root mean square (RMS), or zero-crossing), the ratio between the magnitude difference at the beginning and the ending of a decay period (i.e., decrease slope), and the percentage of frames showing less energy than the average energy (i.e., low energy).
    \item {\bf Frequency-domain features:} They are represented by the remainder indexes, i.e., the rest of features. These features represent the shape of the spectrum and the harmonic structure of sounds such as the fundamental frequency of the audio signal (i.e., pitch), the proportion of frequencies that are not multiple of the fundamental frequency (i.e., inharmonicity), the spectral representation based on the 12 equal-tempered pitches of Western music (i.e., chromagrams), the spectral statistical moments (i.e., centroid, the first one; spread, the second one; skewness, the third one), the ratio between the geometric mean and the arithmetic mean (i.e., flatness), the spectral changes between two successive frames (i.e., flux), and the estimation of the amount of high frequency (i.e., roll-off).
\end{itemize}
For further information, the complete database can be found at \url{https://metacreation.net/emo-soundscapes/}.

\subsection{Multiple Linear Regression Model}\label{sec_MLR}
Let us consider a set of $R$ variables  ${\bf x}=[x_1,...,x_R]^{\top}$ (input vector) and a related variable $y$ (output). In several real-world applications, we observe a dataset of $N$ pairs  $\{{\bf x}_n,y_n\}_{n=1}^N$. In this work, we consider the case that $R\leq N$. The relationship between inputs and  outputs is then studied. 
A linear observation model is usually used,
\begin{align}\label{aquiM0}
    y_n&=\beta_0+\beta_1 x_{n,1}+\beta_2 x_{n,2}+...\beta_R x_{n,R}+ \epsilon_n, \\
      y_n&=\beta_0+ {\bf x}_n^{\top}{\bm \beta}+ \epsilon_n,  \nonumber
\end{align}
where ${\bm \beta}=[\beta_0,\beta_1,...,\beta_R]^{\top}$ is a vector of coefficients and $ \epsilon_n$ is a Gaussian noise with zero mean and variance $\sigma_\epsilon^2$, i.e.,  $ \epsilon_n \sim \mathcal{N}(\epsilon|0,\sigma_\epsilon^2)$.
Defining the vectors ${\bf y}=[y_1,...,y_N]^{\top}$ and ${\bm \epsilon}=[\epsilon_1,...,\epsilon_N]^{\top}$,  and the $N\times (R+1)$ {\it design matrix} defined as  
$$
{\bf X}=
\begin{bmatrix}
1 & {\bf x}_1^{\top} \\
\vdots  &  \vdots \\
1 & {\bf x}_N^{\top}
\end{bmatrix},
$$
the previous model can be rewritten in the following way,
\begin{align}\label{aquiM}
      {\bf y}={\bf X}{\bm \beta}+ {\bm \epsilon}. 
\end{align}
The least squares (LS) estimator (which coincides with the maximum likelihood estimator, in this setting) is
\begin{align}
      \widehat{{\bm \beta}}=\left({\bf X}^{\top}{\bf X}\right)^{-1} {\bf X}^{\top} {\bf y}.
\end{align}
Hence, the vector of output predictions according to the model is $\widehat{{\bf y}}={\bf X}\widehat{{\bm \beta}}={\bf X}\left({\bf X}^{\top}{\bf X}\right)^{-1} {\bf X}^{\top} {\bf y}$, and the error vector is 
\begin{align}
      \widehat{{\bf e}}={\bf y}-\widehat{{\bf y}}=\left({\bf I}-{\bf X}\left({\bf X}^{\top}{\bf X}\right)^{-1} {\bf X}^{\top}\right) {\bf y},
\end{align}
where $ \widehat{{\bf e}}=[\widehat{e}_1,...,\widehat{e}_N]^{\top}$, and ${\bf I}$ is a $N\times N$ unit (diagonal) matrix. The mean absolute error (MAE) and the mean squared error (MSE) - in one specific  realization\footnote{The MAE and MSE are theoretically defined as the expectation of the (abosolute or squared) error over different realizations of the data ${\bf y}^{(\ell)}$ with $\ell=1,2,3,$..., where the index $\ell$ denotes a different realization. Here, we consider a fixed vector of data ${\bf y}$ (i.e., the observed data) and compute  the error vector in one realization $\widehat{{\bf e}}={\bf y}-\widehat{{\bf y}}$. Moreover, we average the absolute value (or the squared values) of error components in order to obtain a unique error value.   }  - are defined as $\texttt{MAE}=\frac{1}{N}\sum_{n=1}^N |\widehat{e}_n|$ and $\texttt{MSE}=\frac{1}{N}\sum_{n=1}^N \widehat{e}_n^2$, respectively. We also denote as $y_1$, $y_2$ the output 1 (arousal) and the output 2 (valence), respectively.

\section{Robust variable selection analysis}\label{sec_varanalysis}
Variable selection is one of the most important task in machine learning, signal processing and statistics. The main goal of a variable selection approach is to remove the redundancy contained in the data, ideally without any loss of information or, at least, without incurring in a sensible loss of information.
\newline
Variable selection methods are conceptually formed by two main parts. The first stage consists in {\it ranking} the variables for their importance, measured with some suitable criterion. In a second stage, based on the previous ranking, a  selection of {\it the number of relevant variables} is performed. This last part can be considered a dimension reduction step, and it is also strongly connected to the model selection problem in nested models  (in this case, the model selection problem is often called {\it order} selection) \cite{Stoica04digital,Stoica04SPM,llorente2020review}. 
\newline
Generally, in the literature, these two stages are jointly presented within a unique  technique, including the second part as a {\it stopping rule} in the ranking procedure (e.g., using a {\it threshold  value} and stopping the rank at some position once the threshold value is reached)  \cite{VarSelRev_paper}. Here, we describe separately these two stages: the ranking methods in the next subsection and the selection of the number of features in Section \ref{SectNumRelVar}. Hence, we can combine different ranking schemes with different procedures for selecting the number of variables.
\newline
In this work, we describe five different ranking methods (RMs). The first four ranking procedures are based, in a different way, on the prediction error. To the best of our knowledge, the procedures RM3 and RM4 described below present also some degree of novelty. They allow us to perform a more robust analysis, as discussed in Section \ref{sec_results1}. As an additional final check on the obtained results, we also apply a classical ranking method based on p-values \cite{efroymson1960multiple,hocking1976biometrics}.
\newline
In Section \ref{BSSect}, we also describe the best sequence search (keeping fixed the number $M$ of variables in the sequence) and an alternating optimization technique for obtaining the optimal sequence. 
Furthermore, we introduce a target density  based on the prediction error, and employ a Gibbs sampling scheme for drawing samples from it.
This analysis allows to have a complete view of the importance of the variables. 

\subsection{Ranking methods}
In this section, we briefly describe the ranking methods (RMs) that we have applied to our dataset. Some of them are well-known techniques, whereas others contain some degree of novelty \cite{VarSelRev_paper}. We list them below.
\newline
\newline
 {\bf RM1 - Forward Selection (FS):} {\it adding variables ``forward'' minimizing the error}. The method starts searching for the most significant univariable model (in terms of the error in prediction), i.e., considering the linear regression model in Eq. \eqref{aquiM0} with only one component (namely, one column in the matrix ${\bf X}$ in Eq. \eqref{aquiM}). We repeat then considering a model with two variables (re-estimating the model for each pair), including (and keeping) the previously selected variable.
We iterate the procedure until considering a complete model of $R$ variables. This procedure provides a sequence of included variables that will be the final ranking. 
\newline
{\bf RM2 - Backward Elimination (BE):}  {\it removing variables ``backward'' minimizing the error.} The method starts considering the complete model. Then, we remove the most insignificant
variable in terms of the error in prediction, considering models with $R-1$ variables (clearly, removing a different variable and we re-estimate the coefficients for each model). We repeat the procedure considering always smaller models and removing one variable at each iteration. The procedure provides an inverse ranking where the first variable is the worst one and the last is the best one. 
\newline
 {\bf RM3:} {\it removing variables maximizing the error.}
The method starts again considering the complete model. Then, we remove one variable and compute the error in prediction. Hence, we select {\it the best variable}, i.e., which (when removed) produces the higher increase of the error in prediction. This variable will be the first in our ranking (the most relevant variable). We repeat the procedure considering the rest of variables.
\newline
 {\bf RM4:} {\it adding variables maximizing the error.} Here, we create a sequence of variables from the worst to the best variable (i.e., increasing  their relevance),
starting with a univariable model as in RM1, but selecting the worst variable (i.e., the variable which maximizes the prediction error). Then, we consider a model with two variables (keeping the previous select one) and select the second variable which maximizes the prediction error. We repeat the procedure, obtaining a final inverse ranking of the variable, i.e., the last one will be the most relevant variable.
\newline 
 {\bf RM5 - based on the correlation coefficient:} We compute the Pearson correlation coefficients between one single variable and the output $y$. Then we   rank the features in decreasing order according to the module of correlation coefficients. This procedure is  similar to the often so-called {\it univariable selection} \cite{VarSelRev_paper}.
\newline
\newline
The joint use of these different ranking procedures allows to perform a robust analysis, obtaining a more complete view of our variable selection problem. Indeed, some ranking methods, although yield sequences far from the smallest possible error, detect relevant variables that appear also by the Gibbs sampling analysis (described below). More specifically, although we will see that RM1 and RM2 provide the best performance in terms of prediction error, but
the results of RM3, RM4 and  RM5 reveal other important aspects shown by the rest of our analyses below.  
Moreover, for completing our view, we will also apply a classical ranking method  based on p-values \cite{efroymson1960multiple,hocking1976biometrics}, and show the results in Table \ref{Tabla3}.
 

\subsection{Best sequence search}\label{BSSect}

Let us define the vector of $M$ different indices
$$
{\bf v}_M=[k_1,...,k_M], \qquad M\leq R, 
$$
where $k_i\in\{1,2,...,R\}$ but $k_i\neq k_j$ for $i\neq j$. Considering only the $M$ variables in ${\bf v}_M$, we can build a smaller $N\times (M+1)$ design matrix  ${\bf V}_M$, and consider a smaller $(M+1)\times 1$ vector of coefficients  $\widehat{{\bm \beta}}_M=[\widehat{\beta}_{1},...,\widehat{\beta}_{M+1}]^{\top}$, which is obtained as 
\begin{align}
      \widehat{{\bm \beta}}_M=\left({\bf V}_M^{\top}{\bf V}_M\right)^{-1} {\bf V}_M^{\top} {\bf y}.
\end{align}
Moreover, we define the cost function
\begin{align}\label{CostFunct}
 C({\bf v}_M)&=||{\bf y}- \widehat{{\bf y}}||_p^\alpha, \nonumber \\
 &=||{\bf y}- {\bf V}_M \widehat{{\bm \beta}}_M||_p^{\alpha},
\end{align}
where $||\cdot||_p$ is the $L_p$ norm with $p>0$ and $\alpha>0$.
Note that $C({\bf v}_M)$ is defined in the discrete space of $M$ possible different indices. We desire to find the vector of indices such that
\begin{equation}\label{BestSequenceEq}
{\bf v}_M^* =\arg \min_{{\bf v}} C({\bf v}_M).
\end{equation}
Note that an exhaustive search is only possible when $M$ is small (typically it is suitable only for $M\leq 4$). Moreover, a random search in the entire space (as with a simulated annealing approach \cite{LuengoMartino2020}) can be very costly and to reach the global minimum (or a ``good'' local minimum) is very difficult. For this reason, we employ an alternating optimization approach that, at least, ensures a fast convergence to a local minimum. Furthermore, we perform the alternating optimization scheme several times ($500$ runs) with different initializations, and compare the minimum obtained at each run \cite{LuengoMartino2020}. We finally consider the solution $\widehat{{\bf v}}_M$ with the smallest associate cost value $C(\widehat{{\bf v}}_M)$, i.e., $\widehat{{\bf v}}_M$ is our estimator ${\bf v}^*_M$. Below, we describe the alternating optimization method. 
\newline
\newline
{\bf Alternating optimization.} Choose $M< R+1$, a maximum number of iterations $T_{\texttt{iter}} \geq 1$, and start with ${\bf v}_M^{(0)}$.
\newline
\newline
For $t=1,...,T_{\texttt{iter}}$ (or until convergence) repeat: 
\begin{enumerate}
   \item For $j=1,...,M$:
   \begin{enumerate}
    \item Keeping fixed the rest of $M-1$ variable, work only on the $j$-th variable, i.e., given
    $$
     {\bf b}_{j}= [k_1^{(t)},k_2^{(t)},...k_{j-1}^{(t)},k_j,k_{j+1}^{(t-1)}...,k_M^{(t-1)}],
    $$
    find 
    $$
    k_j^*=\arg \min_{{\bf v}} C({\bf b}_j).
    $$
    The optimization above can be solved in an exhaustive way since it is a one-dimensional problem. 
    \item Set $k_j^{(t)}= k_j^*$, and
      $$
     {\bf b}_{j+1}= [k_1^{(t)},k_2^{(t)},...k_{j-1}^{(t)},k_j^{(t)},k_{j+1}, k_{j+2}^{(t-1)}...,k_M^{(t-1)}].
    $$
    \end{enumerate}
    \item Set 
    $$
    {\bf v}_M^{(t)}={\bf b}_{M}, \mbox{ and } {\bf b}_{1}= {\bf v}_M^{(t)}.
    $$
\end{enumerate}


\subsection{A Gibbs sampling approach}

We generalize the optimization scheme considering a Markov Chain Monte Carlo (MCMC) sampling approach. More specifically, we consider a Gibbs sampler which is the counterpart of the alternating optimization in the Monte Carlo sampling world \cite{LuengoMartino2020}. The sampling approach (applied in this context) can provide the probability that each variable is contained in the best subset of $M$ elements. Let us recall the vector of $M$ different elements
$$
{\bf v}_M=[v_1,...,v_M],
$$
where $v_i\in\{1,2,...,R\}$ but $v_i\neq v_j$ for $i\neq j$.
In this section, we consider the target density
$$
p({\bf v}_M)\propto \exp\left(-\eta C({\bf v}_M)\right), \qquad \eta>0,
$$
where $C({\bf v}_M)$ is the cost function previously considered in Eq. \eqref{CostFunct}. The constant $\eta$ can be used and set to provide a tempering effect \cite{llorente2020review,LuengoMartino2020,AutomaticTempering}.  The variables that belong to sequences with smaller errors acquire more value according to $p({\bf v}_M)$. Thus, by drawing samples from $p({\bf v}_M)$, we can obtain the proportion of times that a feature provides a sequence with yields a small error. This is a very important information that helps us to yield a more {\it robust} analysis, in the sense that we can avoid overfitting at this specific set of data. The overfitting can occur performing the variable selection only considering the best sequence, for instance.
Note that this idea can be employed in any problem where a cost function (as function of the parameters of interest) is available.
\newline
\newline
{\bf On the choice of $\eta$.} We can observe that, as $\eta \rightarrow 0$, the density $p({\bf v}_M)$ becomes closer and closer to a delta function around the best sequence ${\bf v}_M^*$ in Eq. \eqref{BestSequenceEq}, which is the global minimum of $C({\bf v}_M)$. As $\eta>0$ grows, more and more local modes appear in $p({\bf v}_M)$. These local modes contain relevant information for our study. As $\eta \rightarrow \infty$, the difference among the values of the modes become smaller and smaller, and $p({\bf v}_M)$ tends to a uniform density in support domain. It is important to remark that there is a range of suitable values  of $\eta$ such that the analysis can be performed. These suitable values are all the values of $\eta$ such that all the possible local modes appear. The interested user can perform some preliminary runs for choosing a proper value of $\eta$.   
\newline
\newline
A Gibbs sampling algorithm is a type of Markov Chain Monte Carlo (MCMC) method for drawing samples from general distribution as $p({\bf v}_M)$ above \cite{MartinoA2RMS,FUSS,MARTINOrecGibbs}. An MCMC algorithm generates a Markov chain with invariant density exactly the target density, that in our case is $p({\bf v}_M)$.
A Gibbs sampler works at each step in a one dimensional space \cite{LuengoMartino2020}, simplifying the multivariate sampling problem drawing from simpler one-dimensional densities. Before describing the Gibbs sampling method, we have to recall that  the $j$-th {\it full-conditional density} is
\begin{align}
 & p_j(v_j|v_1,...,v_{j-1},v_{j+1},...,v_M) \nonumber \\
& \propto p({\bf v}_M)=p(v_1,...,v_{j-1},v_j,v_{j+1},...,v_M),
\end{align}
where all the variables are fixed with the exception of $v_j$, and  the normalizing constant is $p(v_1,...,v_{j-1},v_{j+1},...,v_M)$ that does not depend on $v_j$. For simplicity, we use the more compact notation
$$
p_j(v_j|v_1,...,v_{j-1},v_{j+1},...,v_M)=p_j(v_j|v_{1:j-1},v_{j+1:M}),
$$
for denoting the $j$-th full-conditional density. A detailed description of the Gibbs sampling algorithm is given below.
\newline
\newline
{\bf Gibbs sampler.}  Choose $M< R$, a maximum number of iterations $T_{\texttt{iter}} \geq 1$, and start with ${\bf v}_M^{(0)}$.
\newline
\newline
For $t=1, \ldots, T_{\texttt{iter}}$:
\begin{enumerate}
    \item For $j=1, \ldots, M$:
    \begin{enumerate}
    \item  Draw $v_{j}^{(t)} \sim p_j\left(v_{j} \mid v_{1: j-1}^{(t)}, v_{j+1: M}^{(t-1)}\right)$ .
    \end{enumerate}
    \item Set $\mathbf{v}_M^{(t)}=\left[v_{1}^{(t)}, v_{2}^{(t)}, \ldots, v_{M}^{(t)}\right]$.
\end{enumerate}

\subsection{Selection of the number of relevant variables}\label{SectNumRelVar}

Several selection procedures (also denoted as {\it stopping rules} during the ranking process) can be applied.
Clearly, a naive method could be just to set a threshold value (or a percentage) for the prediction error, or by a simple visual inspection of the error curve, i.e., the so-called {\it elbow method} \cite{Elbow_paper,Stoica04digital}.
In the classical variables selection analysis, practitioners and researchers often consider statistical tests (e.g., F-test and t-test), employed sequentially to decide whether individual variables should be included in the model, and a stopping rule based on p-values
\cite{efroymson1960multiple,hocking1976biometrics}.
\newline
Other approaches rely on the so called {\it information criteria} methods \cite{Stoica04SPM,llorente2020review}, which are based on the following cost function
\begin{align} \label{Caqui}
   C(M)
=\underbrace{-2\log p({\bf y}|\widehat{{\bm \beta}}_M)}_{\mbox{\scriptsize fitting }}+\underbrace{ 2 \xi M}_{\mbox{\scriptsize penalization}},
\end{align}
where $\xi >0 $ is a constant that specifies the criterion. The first term is a fitting term (based on the maximum likelihood value), whereas the second term is a penalty for the  model complexity.  The expression \eqref{Caqui} encompasses several well-known information criteria proposed in the literature and shown in Table \ref{TablaIC}, which differ for the choice of $\eta$ \cite{Stoica04SPM,llorente2020review}. 

\begin{table}[!h]	
	\caption{Different information criterion for model selection ($N$ number of data).}\label{TablaIC}
	\vspace{-0.2cm}
	\footnotesize
	\begin{center}
		\begin{tabular}{|c|c|} 
			\hline 
			{\bf Criterion} & {\bf Choice of } $\xi$   \\ 
			\hline 
			\hline 
			Bayesian-Schwarz information criterion (BIC) \cite{schwarz1978estimating} &  $(\log N)/2$ \\
			\hline 
			Akaike information criterion  (AIC) \cite{Spiegelhalter02} &  $1$ \\
			\hline 
			Hannan-Quinn information criterion (HQIC)  \cite{Hannan79}&  $\log(\log(N))$ \\
			\hline 
		\end{tabular}
	\end{center}
\end{table}


%
%



\section{Results for the output 1 - Arousal}\label{sec_results1}
In this section, we describe the results obtained for the first output (arousal). A more complete discussion is provided in the  \texttt{Supplementary material}.
\newline
\newline
To carry out the analysis and selection of variables with which we propose our final linear model for the prediction of the arousal, we will follow the next steps:
\begin{enumerate}
\item First of all, we apply five ranking techniques (from RM1 to RM5) and analyze the variables that appear classified as the most significant (Section \ref{sec_results1_rank}).
\item We apply the best sequence search (Section \ref{sec_results1_best}) and the Gibbs sampling analysis (Section \ref{sec_results1_Gibbs}) and analyze the variables that appear as the most relevant according to these methods that start from a fixed number of variables. We compare and discuss the obtained results with the results  previously provided by the ranking techniques.
\item We summarize all the previous results by classifying the variables into three levels according to their level of global relevance (Section \ref{sec_results1_summary}).
\item We propose the linear model for the prediction of the arousal including only the most relevant variables according to our analysis and we evaluate their prediction error (Section \ref{sec_results1_model}).
\end{enumerate}
Note that these same steps are briefly replicated for the second output (valence) throughout Section \ref{sec_results2}.

\subsection{Results of the Ranking Methods}\label{sec_results1_rank}
\begin{figure*}[!h]
\centerline{
	\subfloat[MAE  versus $M$, ordering the variables. ]{\includegraphics[width=10cm]{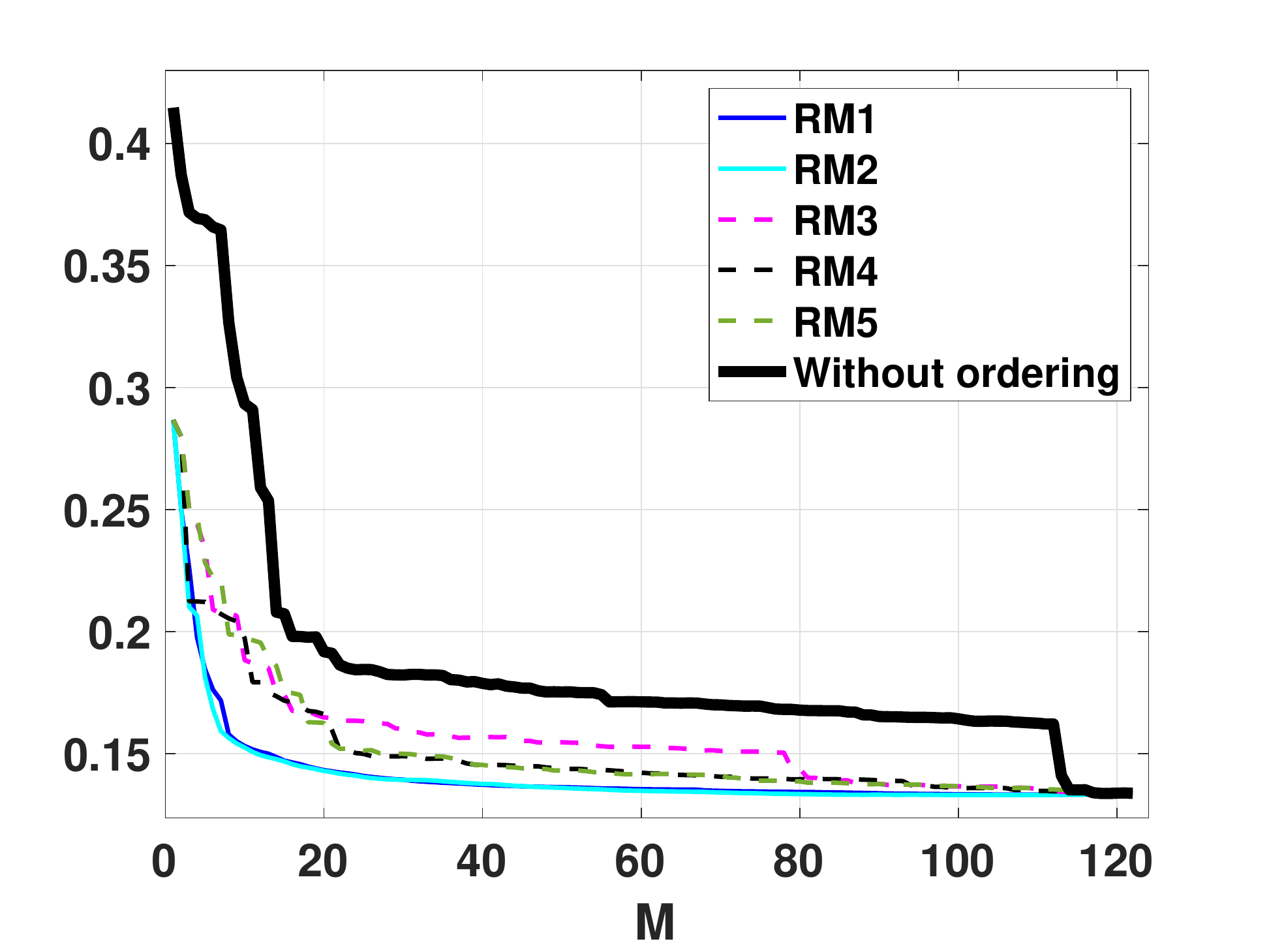}}
	\subfloat[The same of Figure (a) but in log-log scale. ]{\includegraphics[width=10cm]{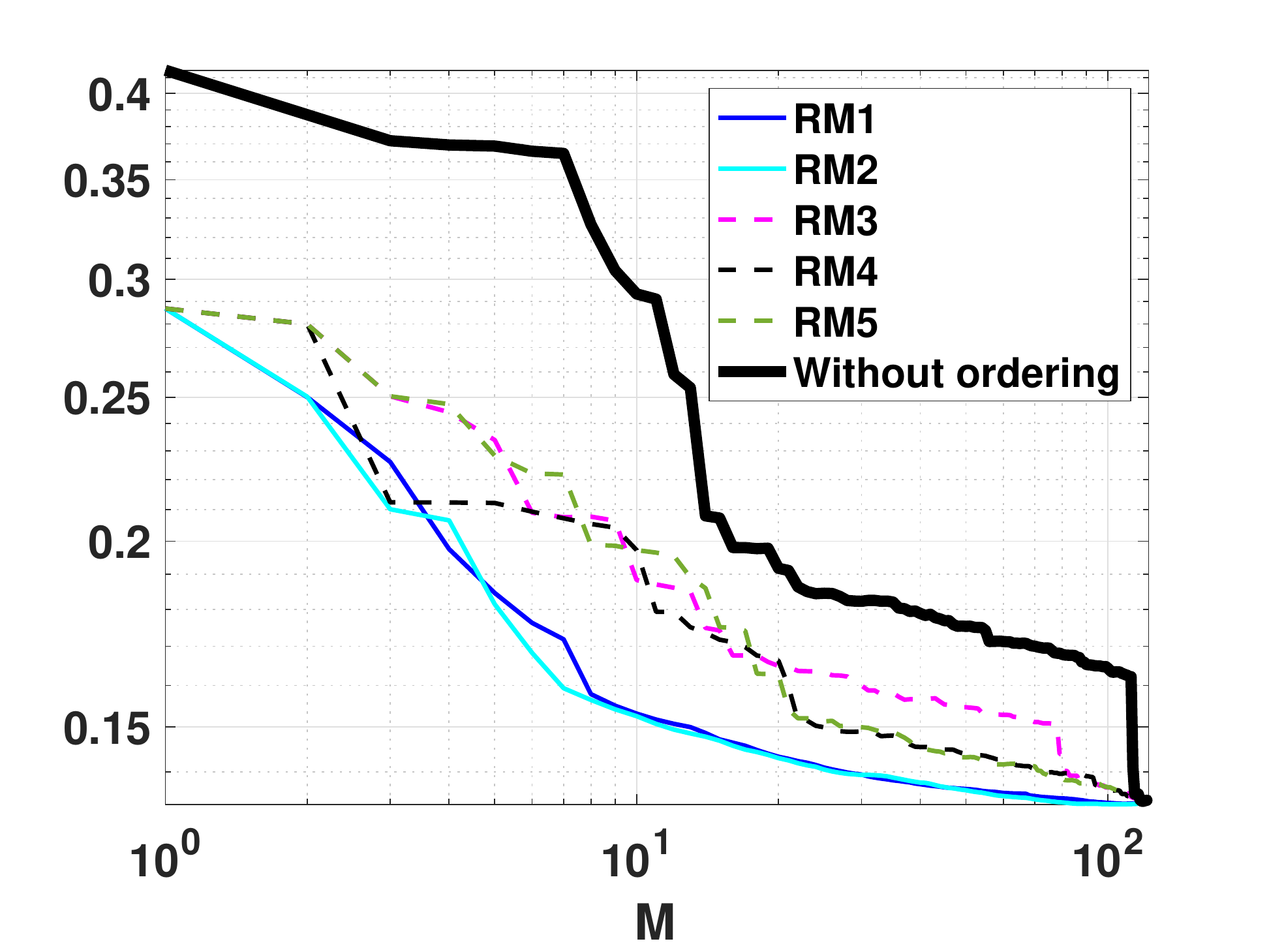}}	
		}		
	\caption{{\bf (Output 1)} MAE versus $M$ obtained ordering the variables according to the different rankings. At each $M$, we consider the first $M$ variables in each ranking and compute the MAE. Clearly, when $M=122$ (i.e., we are using all the variables) all the curves reach the same point. The black solid line corresponds to the MAE curves without ordering the variables. }
	\label{fig1Ranking}
\end{figure*}

Figure \ref{fig1Ranking} shows the MAE in the estimation of the first output considering models with $M\leq 122$ variables. The variables are ordered according to the different ranking criteria. At each $M$, we consider the first $M$ variables in each ranking and compute the MAE. Clearly, when $M=122$ (i.e., we are using all the variables) all the curves reach the same point. The black solid line corresponds to the MAE curves without ordering the variables (i.e., keeping the order in the data matrix).  Note that, even in this curve with  unordered variables, we can observe the importance of the variables  ``12-th chromagram standard deviation'' (indexed as 112),  ``loudness mean'' (indexed as 113) and ``loudness standard deviation'' (indexed as 114).  Indeed, there is a relevant drop in MAE at the variable 112, the decrease continues with the variable 113, and the derivative seems to be null after the variable 114. Moreover, even in this curve with unordered variables,  we can observe in Figure \ref{fig1Ranking}(a) that there is already an {\it elbow} within the interval between the 15-th variable and the 20-th variable  \cite{Elbow_paper}.
\newline
The cyan and blue solid lines  correspond to RM1 and RM2 which provides the best results in terms of MAE. Namely, RM1 and RM2 provide two orders of variables which produce the fastest decays in terms of MAE. Figure \ref{fig1Ranking}(b) provides  the same information of Figure  \ref{fig1Ranking}(a)  but in log-log-scale. Both curves, corresponding to RM1 and RM2, seem to present a clear {\it elbow} between the 7-th and 8-th ordered variables. 
\newline
The curves corresponding to RM3, RM4, RM5 are depicted with dashed lines. 
Although these rankings do not achieve the best results in this figure, they provide interesting information regarding the importance of the variables, as we discussed below. For instance, note that the error with RM4 has a big drop (reaching the best performance given by RM2) when the third variable is considered, which is feature 1. This feature appears in 8-th position of RM5 but is not considered relevant by RM1, RM2, RM3 and by the best sequence search, as we will see later on. Moreover, the Gibbs analysis confirms its relevance (as we will see below).
\newline
\newline
The rankings of the variables obtained by RM1, RM2, RM3, RM4, RM5, from the first one to the 20-th one are given in Table \ref{Tabla1}. We highlight with boxes the variables that  are within the most important twenty variables in all the ranking methods; these variables are five and are labeled as:
\begin{align*}
& 113 \mbox{ (``loudness mean'')},  \nonumber\\ 
&114 \mbox{  (``loudness standard deviation'')}, \nonumber\\
&14 \mbox{ (``spread mean'')}, \nonumber\\
& 13 \mbox{ (``centroid standard deviation'')}, \\
&\mbox{ and } 3 \mbox{ (``decrease slope mean'')}, 
\end{align*}
although, variable 3 is never contained within the most important ten variables within the different rankings. 

\subsection{Results of the best sequence search}\label{sec_results1_best}
\begin{figure*}[!h]
\centerline{
\subfloat[]{\includegraphics[width=8cm]{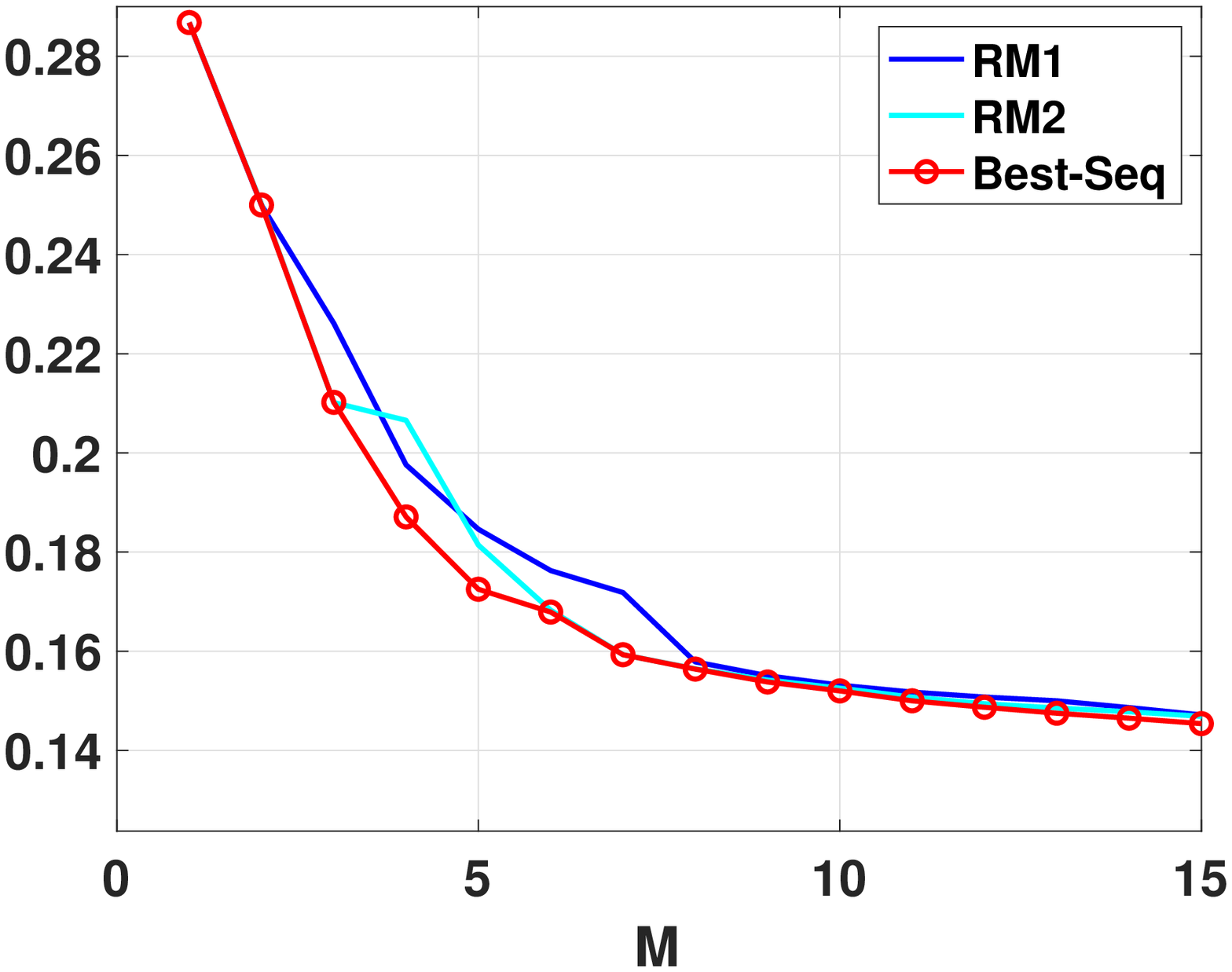}}
\subfloat[]{\includegraphics[width=8cm]{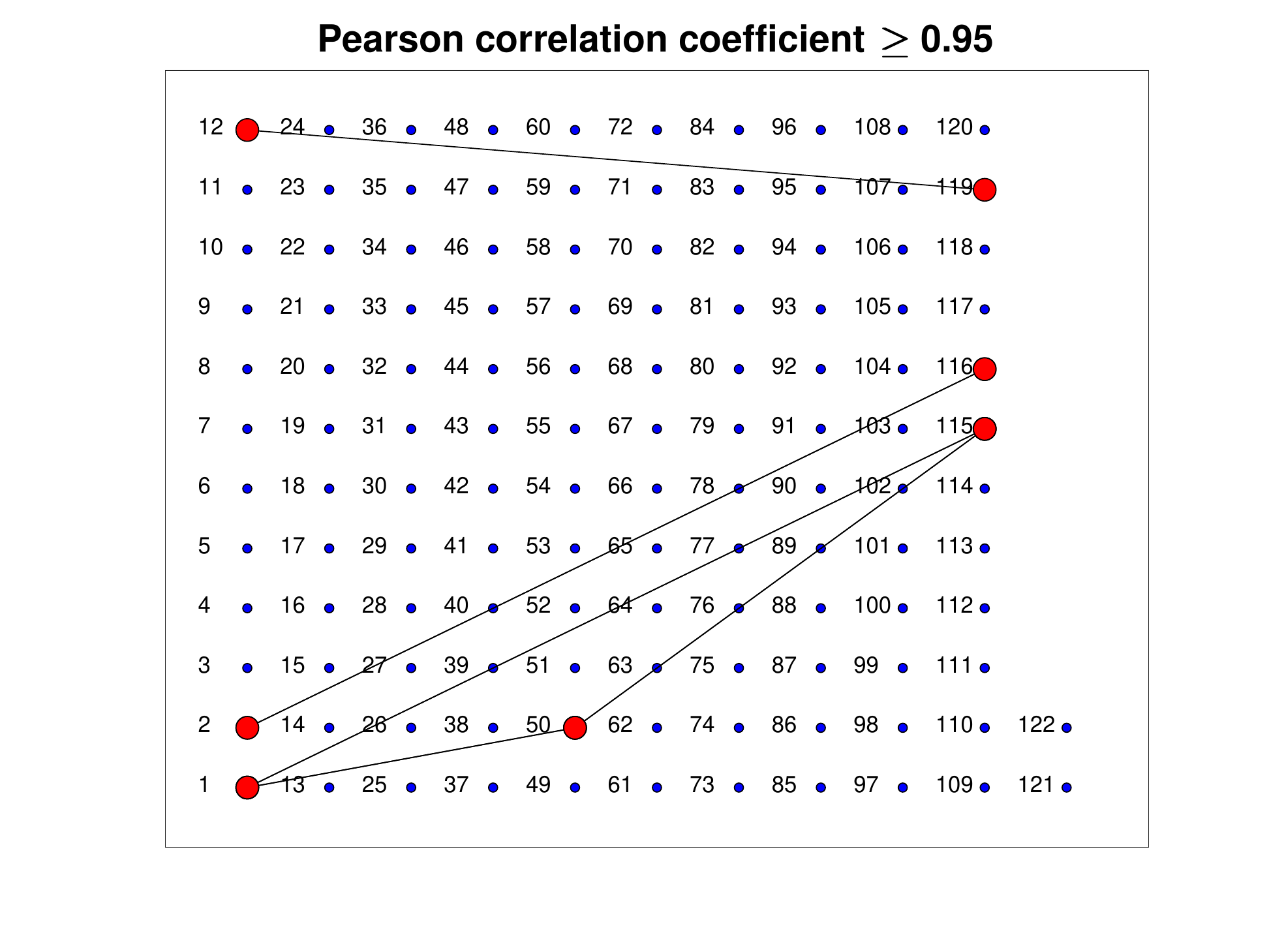}}
}
\vspace{-0.2cm}
	\caption{{\bf (Output 1)} {\bf (a)} MAE versus $M$ obtained ordering the variables according to RM1, RM2 and the best sequence search. {\bf (b)} The connections among the variables with the  Pearson correlation coefficient $\rho$ such that $|\rho|\geq 0.95$.}
	\label{fig2}
\end{figure*}

In Table  \ref{alg:BestSequences2}, we can observe the best sequences for $M=\{1,2,3,4,5,6,7,8,9,10,11,12\}$, obtained with the alternative optimization procedure (after $10^3$ independent runs with different random initializations). See also the decrease of the error in Figure \ref{fig2}(a). All the best sequences are given just in ascending order of the labels. Indeed, unlike with the Gibbs approach,  we cannot discriminate among the variables within a best sequence. 
\newline
In Table  \ref{alg:BestSequences2}, each new entry in the best sequence (as $M$ grows - with respect to the previous shorter sequence), is highlighted with a box. 
We remark especially the best sequence with $M=7$, i.e.,
\begin{align}\label{BestSeq7}
& 4, \mbox{ (``maximum fluctuation'')}, \nonumber \\
&8 \mbox{ (``roll-off mean'')}, \nonumber\\
&14 \mbox{ (``spread mean'')}, \nonumber\\
&56 \mbox{ (``pitch standard deviation'')}, \\
& 113 \mbox{ (``loudness mean'')},  \nonumber\\ 
&114 \mbox{  (``loudness standard deviation'')}, \nonumber\\
&\mbox{ and } 115 \mbox{ (``energy mean'')},  \nonumber\\
& \mbox{(shown here in ascending order of the labels)}. \nonumber
\end{align}
They exactly  coincide with the ranking given by RM2 of the first seven most important variables, i.e.,
\begin{align}\label{BestSeq7_RM2}
113, 14,     8, 114,  115,   56 ,  \mbox{ and }   4,  
\end{align} 
shown here in decreasing order of importance by RM2.

\subsection{Gibbs sampling analysis}\label{sec_results1_Gibbs}

\begin{figure*}[!h]
\centerline{
\subfloat[$M=2$]{\includegraphics[width=9cm]{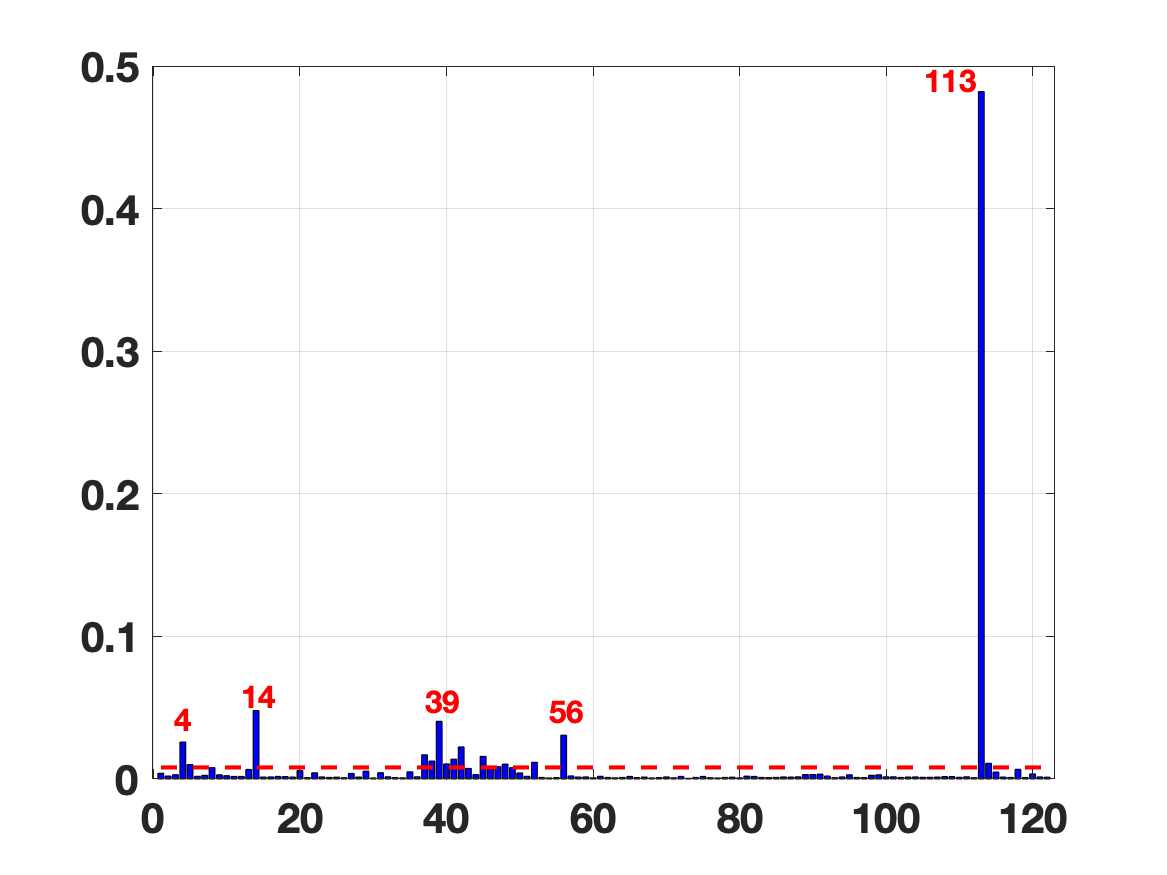}}
		\subfloat[$M=6$]{\includegraphics[width=9cm]{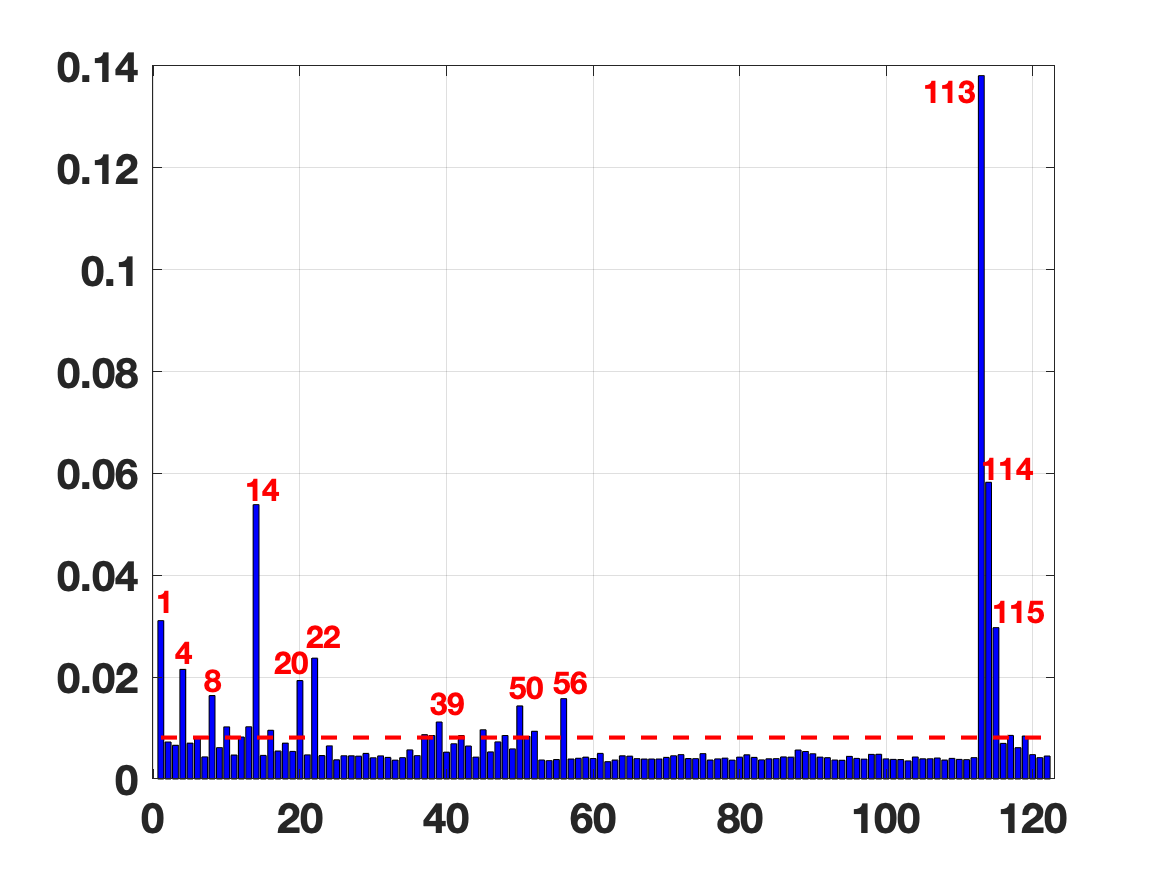}}
}
\centerline{
	\subfloat[$M=10$]{\includegraphics[width=9cm]{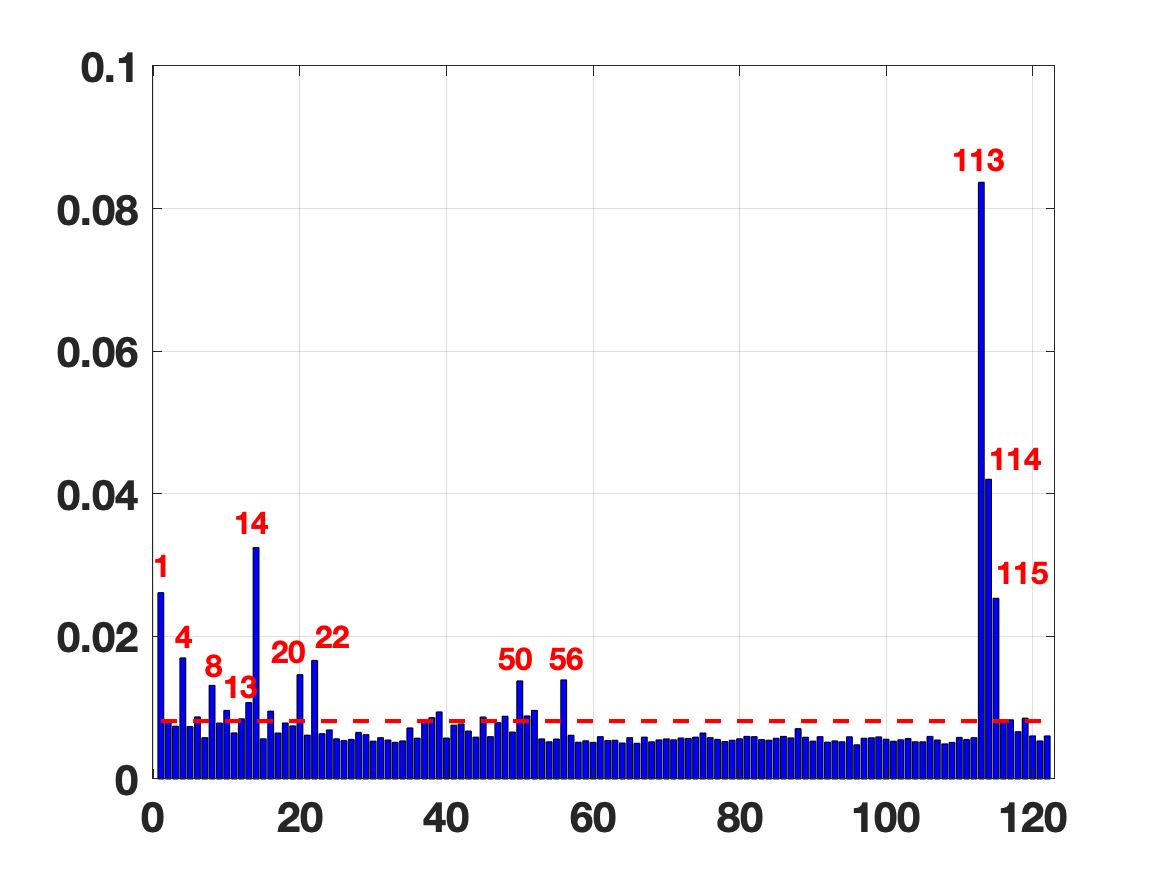}}
		\subfloat[$M=20$]{\includegraphics[width=9cm]{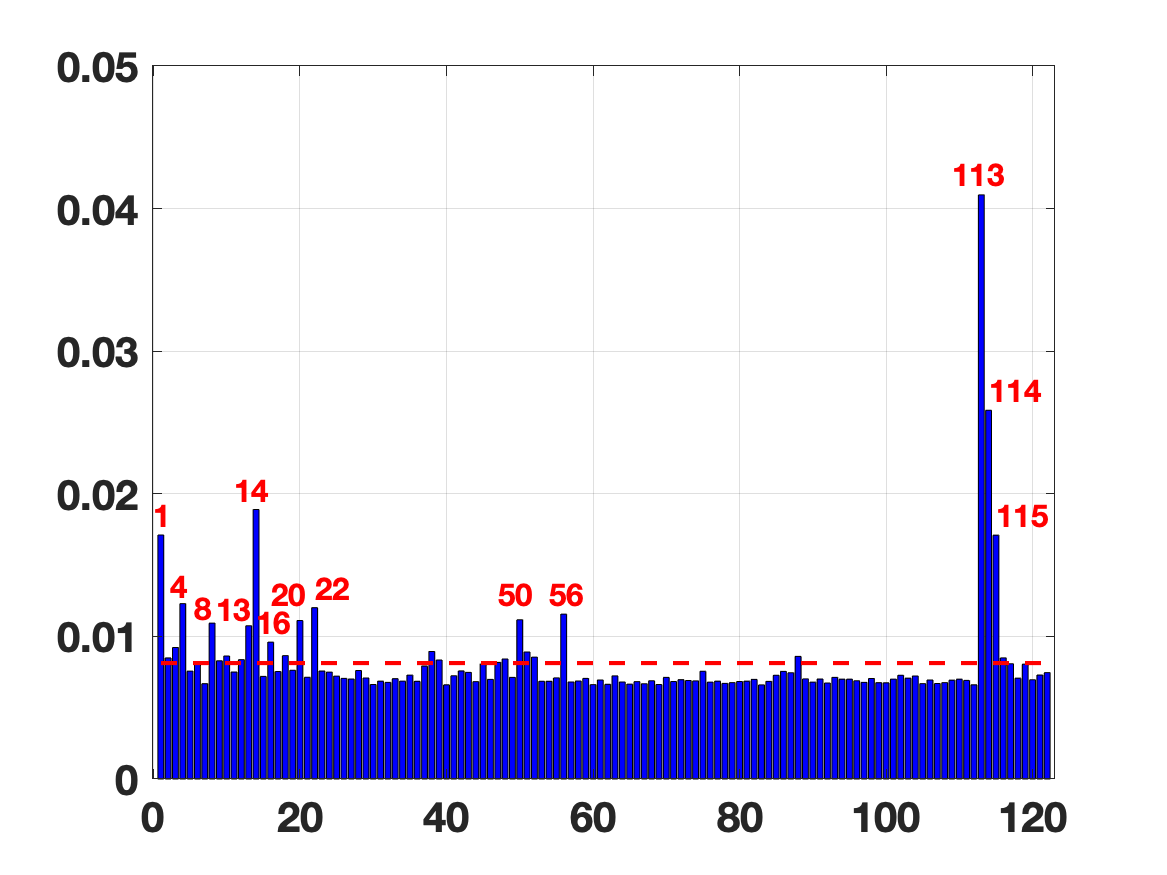}}
		}
	\caption{{\bf (Output 1)} Results in terms of probabilities  obtained by a Gibbs sampling analysis.  The dashed line depicts the equiprobability (uniform) distribution with probability $1/122$.}
	\label{fig1Gibbs}
\end{figure*}

In the Gibbs sampling analysis, the sequences of length $M$, ${\bf v}_M=[v_1,...,v_M]$  (with $v_i\in\{1,2,...,R+1\}$, $v_i\neq v_j$ for $i\neq j$), are weighted according to error $C({\bf v}_M)$ or, more specifically, according to  
\begin{align}
 p({\bf v}_M)\propto \exp\left(-\eta C({\bf v}_M)\right), \qquad \eta>0. 
\end{align}
In our simulation, we set $\eta=100$. Moreover, for defining $C(\cdot)$, we consider the $L_1$ norm (and $\alpha=1$). The variables that belongs to sequences with smaller errors acquire more weight/importance. In some sense, the Gibbs analysis provides the probability that a feature provides a sequence of small error. In Figure \ref{fig1Gibbs}, we show the results of Gibbs sampling analysis for $M\in\{2,6,10,20\}$.  The dashed line depicts the equiprobability (uniform) distribution with probability $1/122$. Clearly, probabilities bigger than $1/122$ denote the most important variables.
\newline
We can observe that the results are coherent with the previous results above. For instance, the variable 113 is clear the most important and also the features $114$, $4$, $115$, $56$, $8$ are quite relevant. The Gibbs analysis also confirms that the feature 39 seems relevant for small $M$ but, as $M$ grows, the importance of this feature disappears.  
\newline
However, by the Gibbs analysis, we can obtain more interesting information. The features $4$ and $56$ are quite relevant even from small values of $M$, confirming also the results of the best sequence search.  The features 14 seems to have a relevance very similar to 114. As we have also previously stated, the Gibbs analysis clearly shows that the variable 115 is the fourth most important feature (as we expect after a care look of the rankings). Surprisingly, the feature 1 seems to be equally relevant than the feature 115: we provide an explanation below. The variable 13 seems also to be some relevance by the Gibbs analysis specially  as $M$ grows. However, as we expected for the previous study, its relevance is moderate.   
\newline
Furthermore, we can also observe the importance of other variables whose relevance was not  clear from the previous analysis above. This is the case of the following features:
\begin{align*}
&1,  \mbox{ (``RMS mean'')},\\
&20,  \mbox{ (``flatness mean'')}, \\
&50, \mbox{ (``flux mean'')},  \\ 
&16,\mbox{ (``skewness mean'')},  \\
&\mbox{ and } 22 \mbox{ (``entropy mean'')}.
\end{align*}
The feature 1 deserves some specific comments (see \texttt{Supplementary Material}).  The feature 20 is in the fourth position of RM3,  in the 10-th position of RM4, and in the 6-th position of RM5. The variable 50 appears also in the fifth position of RM4, and in the 10-th position of RM5. The variable 16 appears in the best sequences for $M\geq 9$ and in 9-th position of RM1. The feature 22  has not  been detected by the previous studies: it does not appear either in the rankings, or in the best sequence search (at least for $M\leq 12$ as in Table \ref{alg:BestSequences2}).

\subsection{Summary for the output 1}\label{sec_results1_summary}
Here, we classify the features into four different classes: {\it very relevant} (Level 1), {\it relevant} (Level 2), and {\it relevant but maybe only for the specific dataset} (Level 3), and the rest of variables belong to the class {\it non-relevant} (Level 4).
\newline
{\bf Level 1.} After all the studies, we can  assert at least $7$ variables are very relevant:
\begin{equation*}
113, 114, 14, 115, 4, 8, \mbox{ and }  56,  \quad \mbox{(ordered by Gibbs analysis),}
\end{equation*}
which are shown in decreasing order of importance considering the Gibbs sampling analysis.
This is also the best sequence for $M=7$, as shown in Table \ref{alg:BestSequences2} and includes also the first $7$ elements  in the ranking obtained by RM2 but with a different order,
\begin{equation*}
113, 14,    8,  114,  115,   56,  \mbox{ and }    4 \quad \mbox{(ordered by RM2)}. 
\end{equation*}
{\bf Level 2.}  Other important variables are 
\begin{equation*}
1, 22, 20, 50, 13, 16, \mbox{ and }  3   \quad \mbox{(ordered by Gibbs analysis)},
\end{equation*}
but the features 1 (``RMS mean'') and 50 (``flux mean'') are highly correlated to the variable 115 (``energy mean''), as shown by Figure \ref{fig2}(b) and as we could intuitively expect. The variable 13 (``centroid standard deviation'') appears in some of the first twelve best sequences. However, the Gibbs analysis reveals (that in terms of robustness) other variables such as 20 (``flatness mean'') and 22 (``entropy mean'') are more or a similar relevance than 13. The feature 20 is particularly important in RM3 (4-th position) and RM5 (6-th position). The  feature 16 (``skewness mean'') does not appears in the best first twenty variables in the rankings, but appears in the best sequences permanently for $M\geq 9$.
In the Gibbs analysis, the feature 16 acquires some relevance as $M$ grows. Finally, the feature 3 (``decrease slope mean'') does not appear in the first twelve best sequences and it does not seems relevant by the Gibbs analysis. However, it appears within the first twenty more relevant variables in {\it all} the ranking methods.\footnote{More surprisingly, for the output 2 - valence -, only three features are included within the first twenty more relevant variables in {\it all} the ranking methods: they are the variables 113, 114 and again 3. Namely the feature 3 (as 113 and 114) is  included within the first twenty more relevant variables in {\it all} the ranking methods for both outputs.   } 
\newline
 {\bf Level 3.} Other possibly important variables, which appear in the best sequence search and in the rankings RM1 and RM2, are
\begin{align*}
&43 \mbox{ (``7th MFCC stand. deviation'')} \\
&\mbox{ and } 107 \mbox{ (``7th chromagram stand. deviation'')}. 
\end{align*}
The variable 43 appears in 8-th position  in RM2 and 13-th position in RM1. Moreover, it appears in the best sequences for $M\geq 8$. The feature 43 appears in 11-th position in RM2 and 14-th position in RM1. Moreover, it appears in the best sequences for $M\geq 10$. On the other hand, the Gibbs analysis does not associate any particular relevance to these variables.

\subsection{Selection of the number of variables and suggested model for the output 1}\label{sec_results1_model}
First of all,  we have applied different information criteria, such as  the  AIC and BIC, shown in Table \ref{TablaIC}  \cite{llorente2020review}.
The more adequate results have been provided the  Bayesian information criterion (BIC) which suggests to use 17 variables whereas AIC suggests the use of 40 variables. We have also tried  the classical analysis based on p-values  which suggests  71 variables \cite{efroymson1960multiple,hocking1976biometrics}.  The results of the corresponding ranking method is given in Table \ref{Tabla3}. The first most relevant $7$ variables are again $113$,   $14$,  $8$, $4$, $56$, $115$ and $114$, i.e., the {\it very relevant} features that we have found after our analysis.
\newline
However, after our exhaustive study, we believe that a more parsimonious model can be suggested.  The most parsimonious LM after all the consideration in our study, is the model which includes at least the seven {\it very relevant} variables (described above),
\begin{equation}
113, 114, 14, 115, 4, 8, \mbox{ and } 56.
\end{equation}
More precisely, the suggested linear model for the output 1 is 
\begin{align}\label{Model1_1}
y_1=&-0.5293 +3.6494 \ x_{113}+1.8080 \ x_{114}+ \nonumber \\
&-1.5534 \ x_{14}  
-3.8491 \ x_{115}+ \\
&+1.5056 \ x_4+ 1.1714 \
x_{8} -0.3450 \ x_{56},\nonumber
\end{align}
obtaining a MAE of 0.1593, MSE of 0.0432 (i.e, RMSE of 0.2078), and $R^2=0.8703$. Considering a Monte Carlo cross-validation  procedure (with $80\%$ of the data in the train-set and the rest of $20\%$ of data in the test-set, chosen randomly  in each $2\cdot 10^4$ independent runs), we obtain  MAE of 0.1611, MSE of 0.0450 (i.e, RMSE of 0.2118), and  $R^2=0.8641$. Namely, we have a very slight increase of MAE and MSE (or a  slight decrease of $R^2$), proving the robustness of our proposed model.

\section{Results for the output 2 - Valence}\label{sec_results2}

In this section, we analyze briefly the results the output 2 of the dataset (valence). The decreases of the error for the RMs are shown in Figure \ref{fig1RankingOUTPUT2}. A complete discussion is provided  in the \texttt{Supplementary Material}. Here, we also point out the  the variables which seem relevant for both outputs (1 and 2) and some features just relevant for output 2.


\subsection{Ranking, best sequences and Gibbs analysis}

The most important features for output 2 are
\begin{align*}
&114 \mbox{  (``loudness standard deviation'')}, \nonumber\\
& 113 \mbox{ (``loudness mean'')},  \nonumber\\ 
&14 \mbox{ (``spread mean'')}, \nonumber\\
&\mbox{ and } 3 \mbox{ (``decrease slope mean'')}. 
\end{align*}
They are also relevant for the output 1 (as we can see in the main body of the work). The variable 114 seems to  increase its relevance with respect the output 1, whereas the variable 3 is much more relevant for the output 2. The importance of these features is confirmed by the Gibbs analysis in Fig. \ref{fig2oputput2}, specially for the feature 14. Indeed, the case of feature 14 is very interesting and reveals also the importance of the Gibbs analysis. The variable 14 does not seem relevant following RM1 and RM2 (which provides the sequences with the smaller errors) and does not appears in the best sequences in Table \ref{alg:BestSequences2_output2}. However, it is the third more important variables for RM3, RM4 and RM5, and it is  the third more relevant variable for the Gibbs analysis (see Fig. \ref{fig2Gibbs}). Furthermore, it acquires more relevance as $M$ grows (see again Fig. \ref{fig2Gibbs}).  The variables 
\begin{align*}
&1 \mbox{ (``RMS mean'')},\\ 
&\mbox{ and } 115 \mbox{ (``energy mean'')},
\end{align*}
(which are highly correlated, also with the feature 50; see the main body of the paper for further details) appear relevant for the second output as well. The variable 1 is contained in the first twenty more important features in RM2, RM4 and RM5. Moreover, the variable 1 appears in the best sequences playing the role of the variable 115, i.e., when the feature 115 does not appear in those sequences. Namely,  due to their correlation in the rankings and in the best sequences, the presence of one of them (1 or 115) avoids the presence of the other one.
The feature 115 appears in the best sequences almost in a stable way for $M\geq 3$.  
The Gibbs analysis confirms the relevance of both 1 and 115, and they seem even more relevant than the variable 3.   
Furthermore, the following variables
\begin{align*}
& 50, \mbox{ (``flux mean'')}, \nonumber \\
& 20, \mbox{ (``flatness mean'')}, \nonumber \\
& 13, \mbox{ (``centroid standard deviation'')}, \nonumber \\
&\mbox{and }  8 \mbox{ (``roll-off mean'')},
\end{align*}
are also important for the output 2. The variable 50 seems relevant bit is highly correlated with the features 1 and 115. 
\newline
The variables above have certain relevance also for the output 1. Now, we discuss some features that seem to have importance only for the output 2 (i.e., valence).
A careful look to the results reveals that the following features
\begin{align*}
& 88 \mbox{ (``inharmonicity standard deviation'')}, \\
&31 \mbox{ (``8th MFCC mean'')}, \\
&40 \mbox{ (``4th MFCC standard deviation'')}, \\
&42 \mbox{ (``6th MFCC standard deviation'')}, \\
& 52 \mbox{ (``Low Energy'')},\\ 
& 79 \mbox{ (``11th chromagram center stand. deviation'')},\\ 
&109 \mbox{ (``9th chromagram stand. deviation'')},
\\ &\mbox{ and } 110 \mbox{ (``10th chromagram standard deviation'')}, 
\end{align*}
are relevant, and appear in the best sequences for the output 2. Moreover, the feature 88 is the second more important in RM1 and appears in the 13-th position of RM3. The importance of 88 is confirmed (and is even more clear) by the Gibbs sampling analysis shown in Figure \ref{fig2oputput2}. The feature 31
seems to be relevant for RM1, RM5 and the Gibbs analysis. Moreover, it appears in the best sequences for $M\geq 11$.
The variable 40 is within the first twenty most important variables of RM1 and RM2. It also appears in the ranking based on p-values in the 10-th position (see Table \ref{Tabla3output2}). Moreover, it starts to appear in the best sequences for $M \geq 10$. The importance of the feature 40 increases as $M$ grows, following the Gibbs analysis. The feature 42 seems to have also the same importance of the variable 40 for the  Gibbs analysis, and appears in the 16-th position of the GR and in the 15-th position of RM1.  The variable 52 takes the 9-th position in RM2, appears in the best sequences for $M \geq 10$ and seems relevant according to the Gibbs analysis.
The feature 79 appears within the first twenty more important variables in RM1, RM2 and RM3. From the Gibbs analysis, it seems clear that its importance grows with $M$. In the best sequences, the feature 79 appears in a stable way for all the best sequences with $M\geq 8$. The feature 110 is contained within the first twenty more important variables in RM1, RM2 and RM3. 
Following  the Gibbs analysis, the feature 109 is similar or more relevant than 110. It also appears in the best sequence with $M=7$ and in the 6-th position of the classical ranking based on p-values. The feature 50 seems relevant by the Gibbs analysis but it is very correlated to 1 and 115. 

\subsection{Selection of the number of variables and suggested model for the output 2} 
From the results, we can notice that the output 2 (valence) is less linear correlated with the variables ${\bf x}$, compared with the first output (arousal). The BIC suggests the use of 22 variables,
whereas the  AIC suggests the use of  68 variables.
The classical p-values method suggests the use of 83 variables. However, all the considerations in our study above, we believe that a more 
parsimonious  model can be proposed. In our opinion, The most parsimonious linear model that we can suggest is the model which includes at least the six {\it very relevant} variables which are  (see the considerations above)
\begin{align*}
&114 \mbox{  (``loudness standard deviation'')}, \nonumber\\
& 113 \mbox{ (``loudness mean'')},  \nonumber\\ 
&14 \mbox{ (``spread mean'')}, \nonumber\\
& 88 \mbox{ (``inharmonicity standard deviation'')}, \\
&115 \mbox{ (``energy mean'')},\\
& 3 \mbox{ (``decrease slope mean'')},
\\
 &\mbox{(ordered  by the Gibbs analysis),}
\end{align*}
and the ten {\it relevant} variables,
\begin{align*}
&8, \mbox{ (``roll-off mean'')}, \\
&20, \mbox{ (``flatness mean'')},\\
&79 \mbox{ (``11th chromagram center stand. deviation'')},\\
&\fbox{4}, \mbox{ (``maximum fluctuation'')},\\
&109 \mbox{ (``9th chromagram stand. deviation'')}, \\ 
&110 \mbox{ (``10th chromagram standard deviation'')}, \\
&40,\mbox{ (``4th MFCC standard deviation'')},\\
&31, \mbox{ (``8th MFCC mean'')},\\
&42,\mbox{ (``6th MFCC standard deviation'')},\\
&\mbox{ and } 52 \mbox{ (``Low Energy'')}, \\
&\mbox{(ordered by the Gibbs analysis)},
\end{align*}
where we have included the feature 4 due to the Gibbs analysis. It appears also in the first twenty positions of RM3, RM4 and RM5: in the 12-th position, in the 18-th position, and in the 15-th position, respectively. 
The variable 72 has been excluded due to the Gibbs analysis, as well.
More precisely, the suggested model for the output 2 is 
\begin{align}
y_2=&0.2831-2.4741 \ x_{114}-1.0919  \ x_{113}+    0.8070\ x_{14}+ \nonumber \\
&0.2538\ x_{88}+2.8482\ x_{115}-0.6448 x_{3}+ \nonumber \\
&-1.4867 \ x_{8} +1.1290\ x_{20} -0.2003\ x_{79}+\\
&-0.7192\ x_4+0.5182\ x_{109}+  0.1642  \ x_{110} +\nonumber  \\
&0.3312\ x_{40} +0.2978 \ x_{31} + 0.4621  \ x_{42}   -0.5342\ x_{52}, \nonumber
\end{align}
obtaining a MAE of 0.2799, MSE of  0.1182 (i.e, an RMSE of 0.3437), and an $R^2=0.6452$. Considering a Monte Carlo cross-validation  procedure (with $80\%$ of the data in the train-set and the rest of $20\%$ of data in the test-set, chosen randomly in each $2\cdot 10^4$ independent runs), we obtain  MAE of 0.2849, MSE of 0.1233 (i.e, RMSE of 0.3509), and  $R^2=0.6311$. As for the output 1, we have a very slight increase of MAE and MSE (and a slight decrease of $R^2$), proving the robustness of our proposed model.


\section{Final discussion and conclusions}\label{sec_conc}

From the previous section, we observe that within the most important features for both outputs, 1 and 2, are
\begin{equation*}
114,113,14,115, \mbox{ and } 3,  \mbox{ (ordered by Gibbs analysis)}.
\end{equation*}
Therefore, we can conclude that the psychoacoustic features 113, 114  (``loudness mean'' and ``loudness standard deviation'',  respectively), the frequency-domain feature 14
 (``spread mean''), and the time-domain feature 115 ( ``energy mean'') are the most relevant variables in our study. They have been included in both suggested models.
 The frequency-domain feature 3 (``decrease slope mean''), although has not been included in the suggested model for the output 1, appears within the first twenty more relevant variables in all the rankings, for both outputs. 
The relevant features reveal the importance of the psychoacoustic indicators in SER. However, time and frequency-domain features have been also included into the suggested models. These results are in line with other studies which also highlight that subjective perception and time-dynamics of the signals (jointly embedded in indicators) lead to better model scores \cite{lionello2020systematic, abri2021comparative}. The valence model provides worse performance than arousal one, even involving more features. This outcome agrees with the literature and it seems to be due to the prevalence of neutral annotations of valence in some soundscape categories \cite{fan2018soundscape}.
\newline
We remark that the suggested LMs are very cheap and parsimonious models (including only 7 variables for arousal and 16 for valence,  over the 122 possible features) and provide quite high $R^2$  coefficients ($R^2=0.8703$ for arousal and $R^2=0.6452$ for valence), and small MSEs (0.045 for arousal and 0.118 for valence),
compared with the results previously obtained in the literature, even using non-linear models and including more variables. Indeed, our results are competitive with respect to the EMO baseline that employs a non-linear SVR and many  more features (exactly 39 variables),  both in arousal (with an MSE of  0.048)  and valence (with an MSE of 0.128)  \cite{fan2017emo}.
In \cite{abri2020predicting},
the authors suggest also linear models with EMO obtaining worse results: specifically  MSE $\approx$ 0.090 for arousal and MSE $\approx$ 0.160 for  valence) using also  more features in their models (exactly 25). Recent studies with EMO have shown that more sophisticated nonlinear models (such as RF) can reach good scores with 15 features for arousal (MSE $\approx$ 0.050) and 14 features for valence (MSE $\approx$ 0.140). 
Finally, other authors using other complex nonlinear models, such as CNNs and data augmentation techniques, obtain slightly better metrics (MSE $\approx$ 0.035 for arousal, and MSE $\approx$ 0.078 for valence), but also including  substantially more variables in their models: from 23 up to 54 features \cite{ntalampiras2020emotional, fan2018soundscape}. All these considerations confirm the quality of our suggested models.
\newline
Due to the exhaustive study that we have performed, we believe that the suggested LMs are robust and allow good prediction in different databases. Thus, the obtained parsimonious models can help the design of SER methods, and its practical applications by remarking the most relevant features.
\newline
As future research lines, we plan to extend our variable selection study (including the proposed Gibbs analysis) for nonlinear models, and then judge if this non-linearity is strongly  required  with the EMO dataset, since the proposed LMs provides already very good performance. Furthermore, we plan to design novel schemes for selecting automatically a reasonable number of variables, when the priority is to obtain the simplest (and hence cheapest) possible model. Indeed, at least with these soundscapes data, the current benchmark techniques often seem to widely overestimate the adequate number of relevant variables.

\section*{{\small  Acknowledgments}}
{\footnotesize 
 The authors acknowledge support by the Agencia Estatal de Investigaci\'on AEI (project SPGRAPH, ref. num. PID2019-105032GB-I00), by  Young Researchers R\&D Project with ref. num. F861 (AUTO-BA-GRAPH)  funded by Community of Madrid and Rey Juan Carlos University and F. Llorente acknowledges support by Spanish government via grant FPU19/00815.
 }

\bibliographystyle{plain}
\bibliography{bibliografia}




\begin{table*}[!h]
\scriptsize
	\centering
	\caption{{\bf Results of the ranking methods - Output 1}}
\makebox[\textwidth][c]{
	  \begin{tabular}{|l||cccccccccccccccccccc|}
	  	\hline
{\bf Meth.} &
{\bf 1} & {\bf 2} &  {\bf 3} & {\bf 4} & {\bf 5}  &  {\bf 6} &  {\bf 7} &   {\bf 8}   & {\bf 9}   & {\bf 10} &  {\bf 11} &  {\bf 12} &   {\bf 13} &  {\bf 14} &   {\bf 15} &{\bf 16 }&  {\bf 17} &   {\bf 18}   & {\bf 19}   & {\bf 20}
 \\	  	
	  	\hline
		\hline
{\bf RM1} &
\fbox{113} & 39 &    \fbox{14} &  8 &    4  &  56 &  115 &   \fbox{114}   & 16   & \fbox{13} &  37 &   77 &   43 &  107 &   55 &    \fbox{3} & 11 &   88   & 38   & 48
 \\
	\hline
{\bf RM2} 
 & \fbox{113} &   \fbox{14} &     8 &  \fbox{114} &  115 &   56 &     4    & 43 &   18&    \fbox{13} &   107 &   55 &   11  &  88    &77 &   38 &     \fbox{3}& 59  &  81  &  78
 \\
	\hline
{\bf RM3} & \fbox{113} &  \fbox{114}&   \fbox{14}  &  20    &18  & 119  &  \fbox{13} &    12   &  9&     8    &21 &    \fbox{3} &   15&    56 &   88   &  4
& 7&    65 &    38 &   39 \\
	\hline
{\bf RM4} &
\fbox{113} &  \fbox{114} &    1 &  115  &  50 &  116  &   2 &   51    &\fbox{14}   & 20 &  119  &  12 &    8&    \fbox{13}  &  9    &21& 117&   18 &   15 &    \fbox{3}
 \\
	\hline
{\bf RM5} & \fbox{113}  &  \fbox{114}  &  \fbox{14} &  116     &2   & 20 &    51    & 1 &  115  &  50 &   \fbox{13} &     9 &     \fbox{3}    &21     &4 &  122 &29 &    56  &  31 &   23 
 \\
	\hline	
	\end{tabular}
}	
	\label{Tabla1}
\end{table*}

\begin{table*}[!h]
	\centering
	\caption{{\bf Best Sequences - Output 1 } }
	      \begin{tabular}{|c||l|l|l|l|l|l|l|l|l|l|l|l|}
		\hline
$M$ & \multicolumn{12}{c|}{{\bf Labels of the features in the best sequence}}  \\
	\hline
	\hline
1 & 113	& & & & & & & & & & & \\
2 &  \fbox{39} & 113 & & & & & & & & & &  	 \\
3   & \fbox{8}   & \fbox{14} &  113 & & & & & & & & &  	 \\
4 &  \fbox{4} &    8    & 14 &   113 	& & & & & & & &  \\
5  & \fbox{10}  &  \fbox{56} &  113  & 114 &  \fbox{115} & & & & & & &   \\
6  & 10  &  \fbox{13} &   56 &  113 &  114 &  115 & & & & & & \\
7  & \fbox{4}  &   \fbox{8} &   \fbox{14}   & 56 &  113 &  114 &  115  & & & & &    \\
8  & 4  &   8  &  14    &\fbox{43}  &  56 &  113 &  114 &  115 & & & &  \\
9  & 4 &    8 &   14    &\fbox{16}&    43  &  56 &  113  & 114 &  115  & & &  \\
10  & 4  &   8  &  14 &   16    &43 &   56 &  \fbox{107}  & 113 &  114  & 115 & &  \\
11  & 4  &   8 &   \fbox{13}    &14   & 16 &   43 &   56 &  107 &  113   &114 &  115 &\\
12 &   4  &   8 &   13   & 14  &  16   & 43 &   \fbox{55}  &  56 &  107  & 113 &  114  & 115 \\
	\hline
	\end{tabular}		
	\label{alg:BestSequences2}
\end{table*}

\begin{table*}[!h]
\scriptsize
	\centering	\caption{{\bf Results of RM based on p-values - Output 1}}
\makebox[\textwidth][c]{
	  \begin{tabular}{|l||cccccccccccccccccccc|}
	  	\hline
{\bf Ranking Method} &
{\bf 1} & {\bf 2} &  {\bf 3} & {\bf 4} & {\bf 5}  &  {\bf 6} &  {\bf 7} &   {\bf 8}   & {\bf 9}   & {\bf 10} &  {\bf 11} &  {\bf 12} &   {\bf 13} &  {\bf 14} &   {\bf 15} &{\bf 16 }&  {\bf 17} &   {\bf 18}   & {\bf 19}   & {\bf 20}
 \\	  	
	  	\hline
		\hline
{\bf RM based on p-values} &
113 &   14 &  8 &    4  &  56 &  115 & 114   & 16   & 43 &  75 & 55 & 13 &  37 &  122 &   2 &  107 & 11 &   40   & 3  & 38  \\
	\hline	
	\end{tabular}
}	
	\label{Tabla3}
\end{table*}

\begin{figure*}[!h]
\centerline{
	\subfloat[MAE  versus $M$, ordering the variables. ]{\includegraphics[width=10cm]{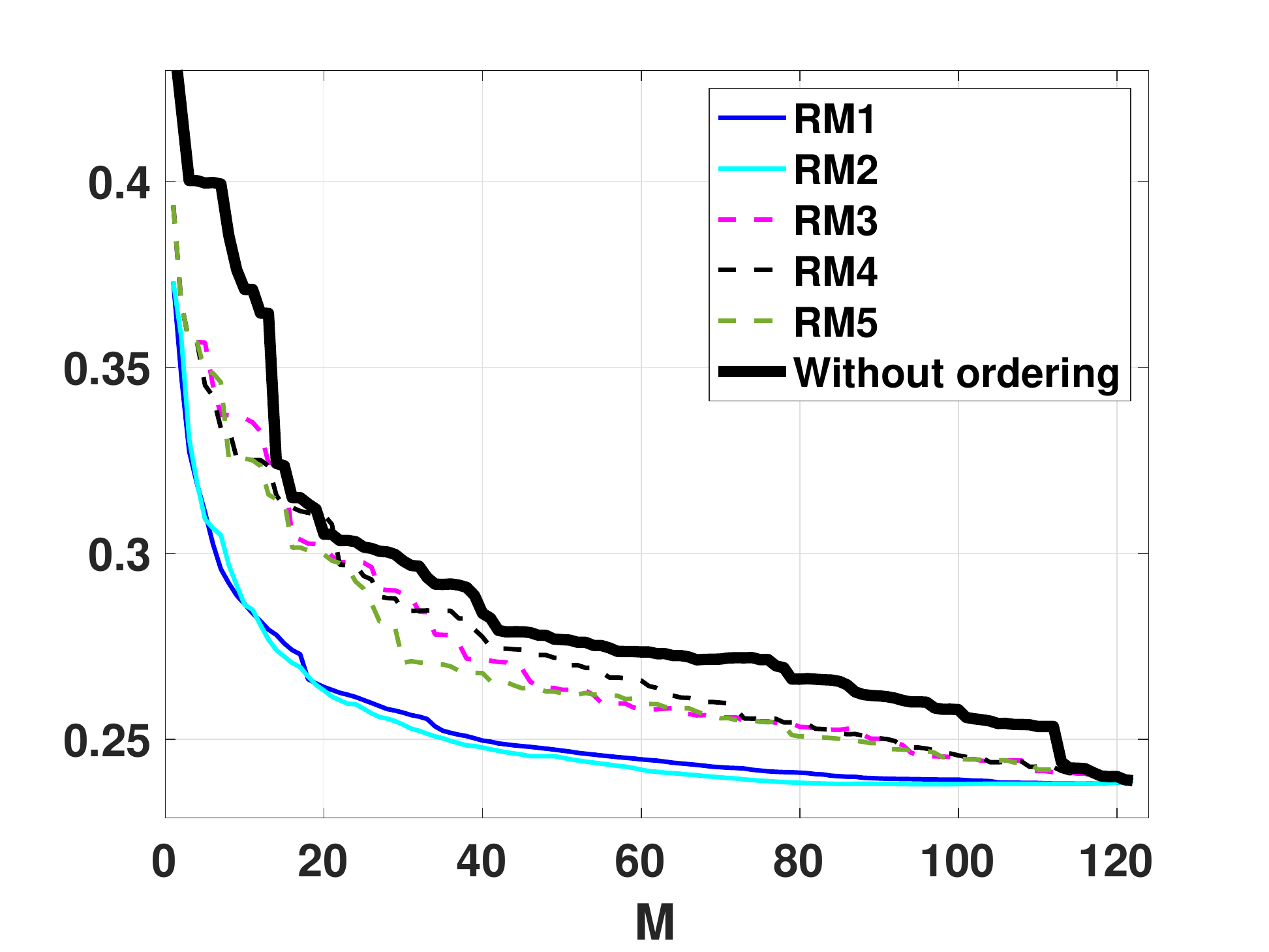}}
	\subfloat[The same of Figure (a) but in log-log scale. ]{\includegraphics[width=10cm]{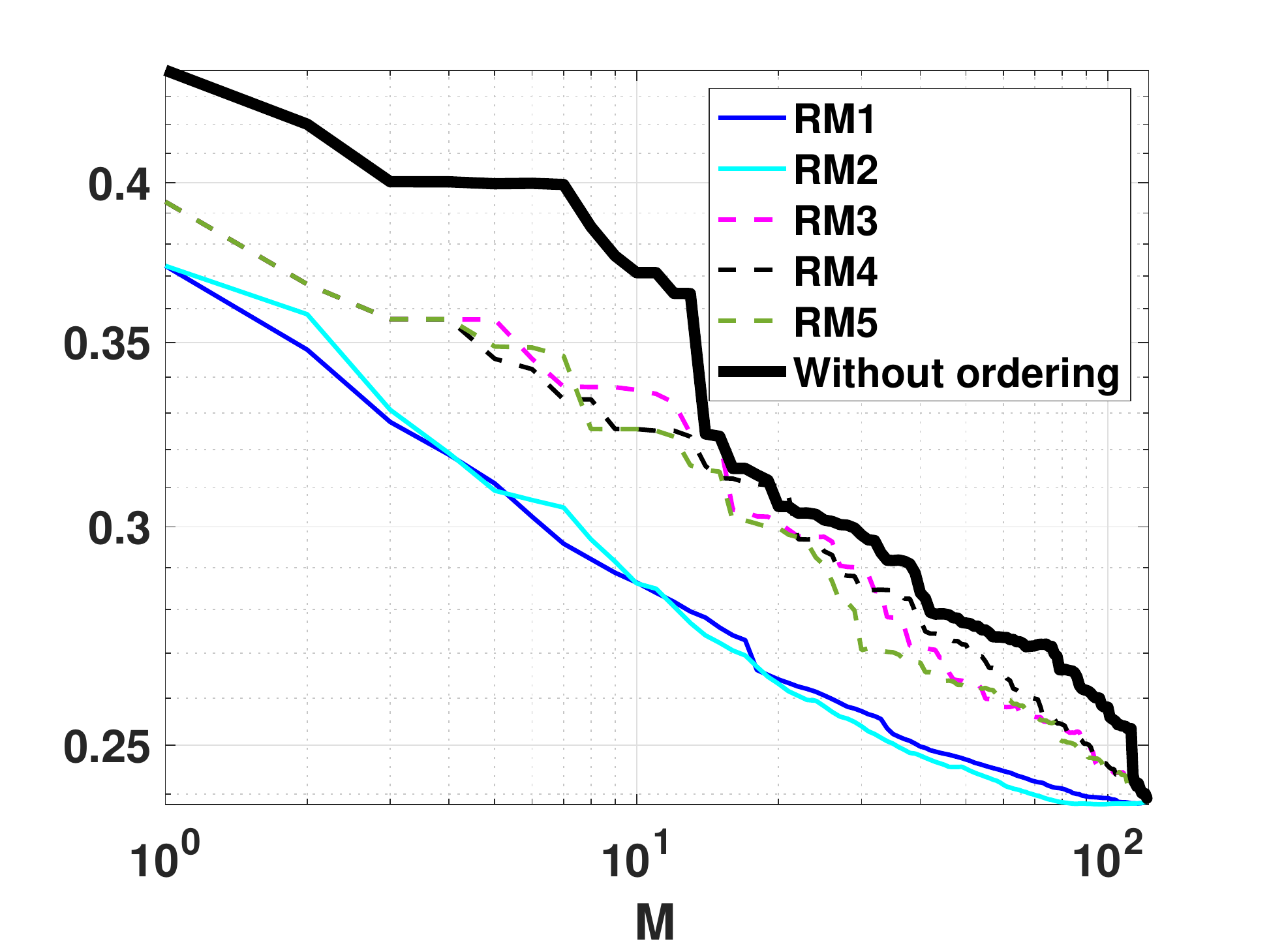}}	
		}
		
	\caption{{\bf (Output 2)} MAE versus $M$ obtained ordering the variables according to the different rankings. At each $M$, we consider the first $M$ variables in each ranking and compute the MAE. Clearly, when $M=122$ (i.e., we are using all the variables) all the curves reach the same point. The black solid line corresponds to the MAE curves without ordering the variables.} 
	\label{fig1RankingOUTPUT2}
\end{figure*}
  
 \begin{figure*}[!h]
\centerline{
\includegraphics[width=9cm]{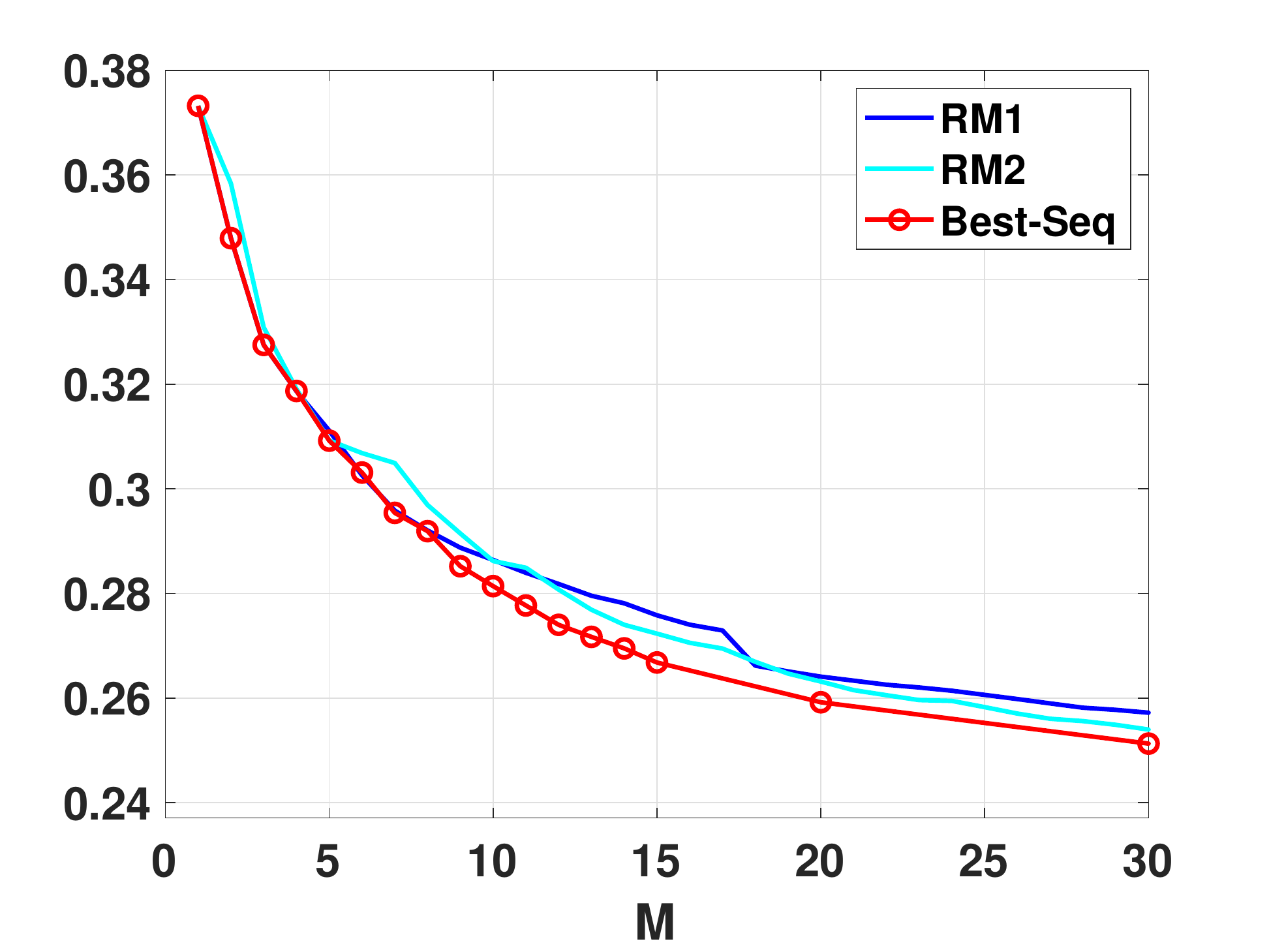}
}
	\caption{{\bf (Output 2)} MAE versus $M$ obtained ordering the variables according to RM1, RM2 and the best sequence search.}
	\label{fig2oputput2}
\end{figure*}

\begin{table*}[!h]
\scriptsize
	\centering
	\caption{{\bf Results of the ranking methods - Output 2}}
\makebox[\textwidth][c]{
	  \begin{tabular}{|l||cccccccccccccccccccc|}
	  	\hline
{\bf Meth.} &
{\bf 1} & {\bf 2} &  {\bf 3} & {\bf 4} & {\bf 5}  &  {\bf 6} &  {\bf 7} &   {\bf 8}   & {\bf 9}   & {\bf 10} &  {\bf 11} &  {\bf 12} &   {\bf 13} &  {\bf 14} &   {\bf 15} &{\bf 16 }&  {\bf 17} &   {\bf 18}   & {\bf 19}   & {\bf 20}
 \\	  	
	  	\hline
		\hline
{\bf RM1} &
\fbox{114}  &  88 &   115   &  \fbox{3} &   42    & 5 &  \fbox{113} &   109 &   64 &   31  &  59 &   15 &    9 &  121 &    13 &    36 &    79 & 75 &    40  & 110
 \\
	\hline
{\bf RM2}  & \fbox{114}  &   1 &  \fbox{113} &   110 &    \fbox{3} &   79 &    94 &    62 &    52 &    72 &    12 &    14 &    40 &    15&    10&   101 &   91 &    92&    29 &    67 
 \\
	\hline
{\bf RM3} & \fbox{113} &  \fbox{114} &   14 &   20 &    18 &     \fbox{3} &   119 &     9 &    13 &    65 &    94 &     4 &    88 &    86  &   79 &   110 &  101 &   121 &    34 &    42 \\
	\hline
{\bf RM4} & \fbox{113} &  \fbox{114} &    14 &   20 &    50 &    51 &     2 &  116 &    1 &   115 &    18 &     9 &    13 &     \fbox{3} &   119 &    19 & 17 &     4 &     12 &    34
 \\
	\hline
{\bf RM5} & \fbox{113} &  \fbox{114} &    14 &   116 &     2 &    20 &    51 &     1 &   115 &    50 &    13 &     9 &     \fbox{3} &    21 &     4   & 122 & 29    & 56   & 31  &  23
 \\
	\hline	
	\end{tabular}
}	
	\label{Tabla1_output2}
\end{table*}

\begin{table*}[!h]
\scriptsize
	\centering
	\caption{{\bf Results of RM based on p-values - Output 2}} \label{Tabla3output2}
\makebox[\textwidth][c]{
	  \begin{tabular}{|l||cccccccccccccccccccc|}
	  	\hline
{\bf Ranking Method} &
{\bf 1} & {\bf 2} &  {\bf 3} & {\bf 4} & {\bf 5}  &  {\bf 6} &  {\bf 7} &   {\bf 8}   & {\bf 9}   & {\bf 10} &  {\bf 11} &  {\bf 12} &   {\bf 13} &  {\bf 14} &   {\bf 15} &{\bf 16 }&  {\bf 17} &   {\bf 18}   & {\bf 19}   & {\bf 20}
 \\	  	
	  	\hline
		\hline
{\bf RM based on p-values} &
114 &   88 &  115 &    3  &  113 & 109 &  5 &
 25 & 65 & 40 & 33 & 37 & 4 & 122 & 68 & 107 & 36 & 2 & 51 & 47
  \\
	\hline	
	\end{tabular}
}	
\end{table*}

\begin{table*}[!h]
	\centering
	\caption{{\bf Best Sequences - Output 2 }} 
	      \begin{tabular}{|c||l|l|l|l|l|l|l|l|l|l|l|l|}
		\hline
$M$ & \multicolumn{12}{c|}{{\bf Labels of the features in the best sequence}}  \\
	\hline
	\hline
1 & 114	& & & & & & & & & & & \\
2 &  \fbox{88}  & 114 & & & & & & & & & &  	 \\
3   & 88 &  114 &  \fbox{115} & & & & & & & & &  	 \\
4 &  \fbox{3}  &  88 &  114 &  115  	& & & & & & & &  \\
5  & 3 &  \fbox{104} &  \fbox{113}  & 114 &  115  & & & & & & &   \\
6  & \fbox{33}  &  \fbox{40}  & 113 &  114 &  115 &  \fbox{119}  & & & & & & \\
7  & \fbox{3}  & \fbox{5} &   \fbox{88}  & \fbox{109}  & 113  & 114 &  115   & & & & &    \\
8  & 3  &   5  &  \fbox{72} &    \fbox{79} &   \fbox{110} &  113 &  114 &  115  & & & &  \\
9  & 3   &  5  &  72 &    79 &   \fbox{88} &   110 &   113 &   114 &  115 & & &  \\
10  & 3 &    5 &   \fbox{40} &   72 &    79 &    88 &   \fbox{112} &   113 &   114 &  115 & &  \\
11  & 3   & \fbox{31} &    40 &   \fbox{52} &    72 &    79 &    \fbox{91} &   \fbox{110} &   113 &   114 &   115  &\\
12 &  \fbox{1} &  3 & 31 &    40 &    52 &    72 &    79 &    \fbox{88} &    91 &   110 &   113 &   114 \\
	\hline
	\end{tabular}		
	\label{alg:BestSequences2_output2}
\end{table*}

\begin{figure*}[!h]
\centerline{
\subfloat[$M=2$]{\includegraphics[width=9cm]{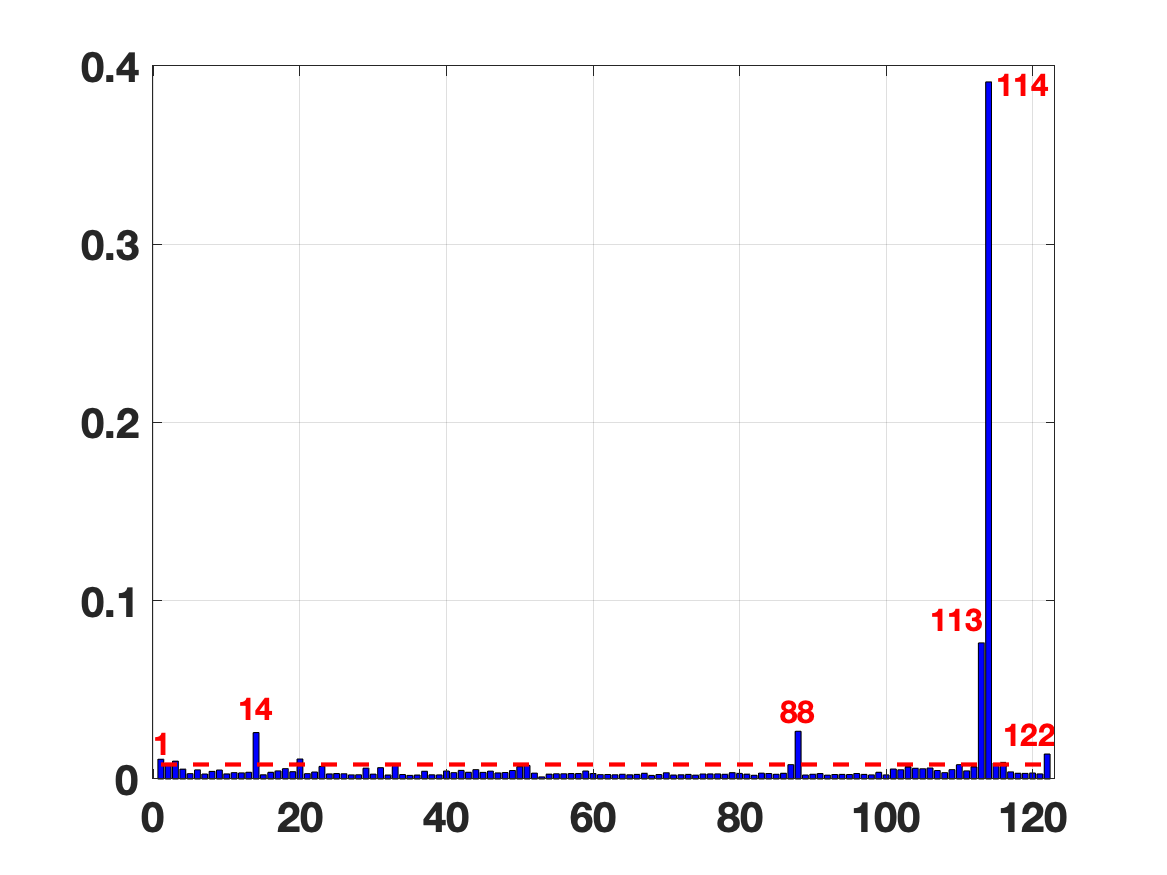}}
		\subfloat[$M=6$]{\includegraphics[width=9cm]{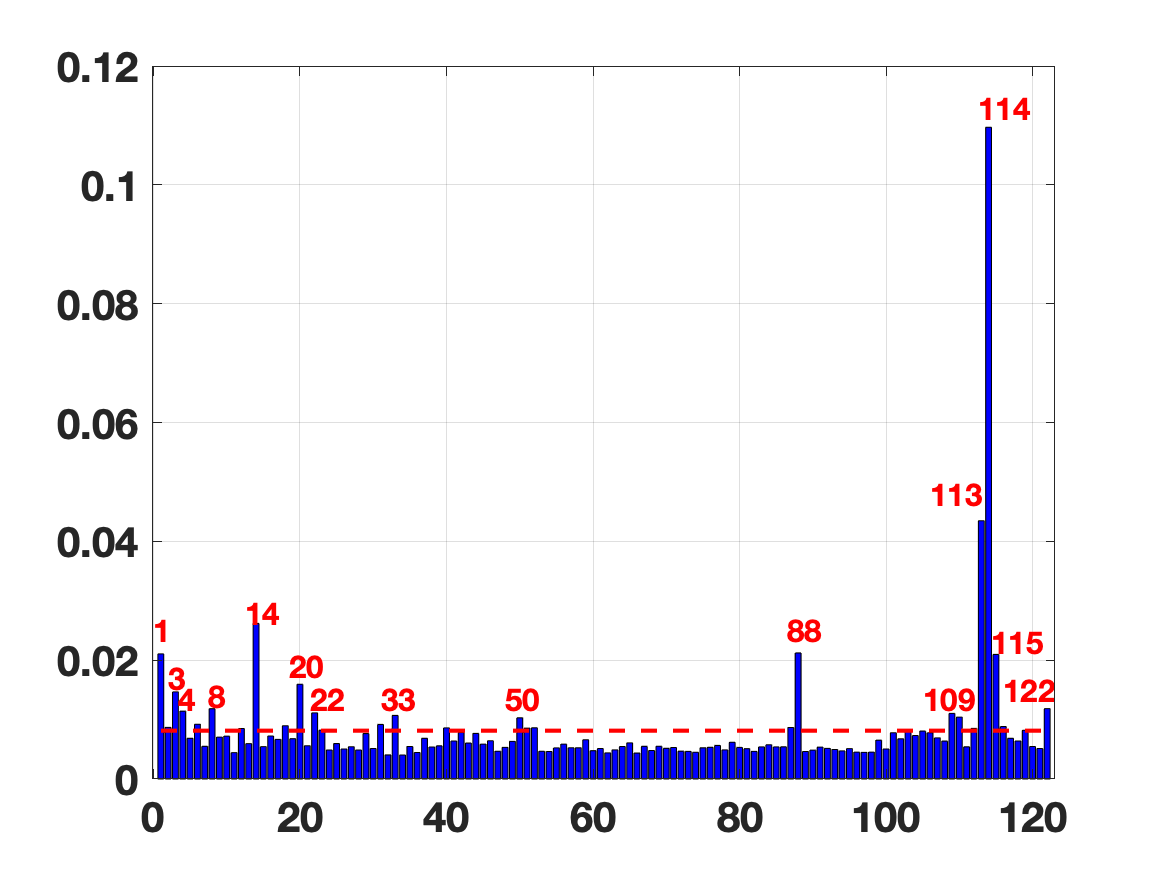}}
}
\centerline{
	\subfloat[$M=10$]{\includegraphics[width=9cm]{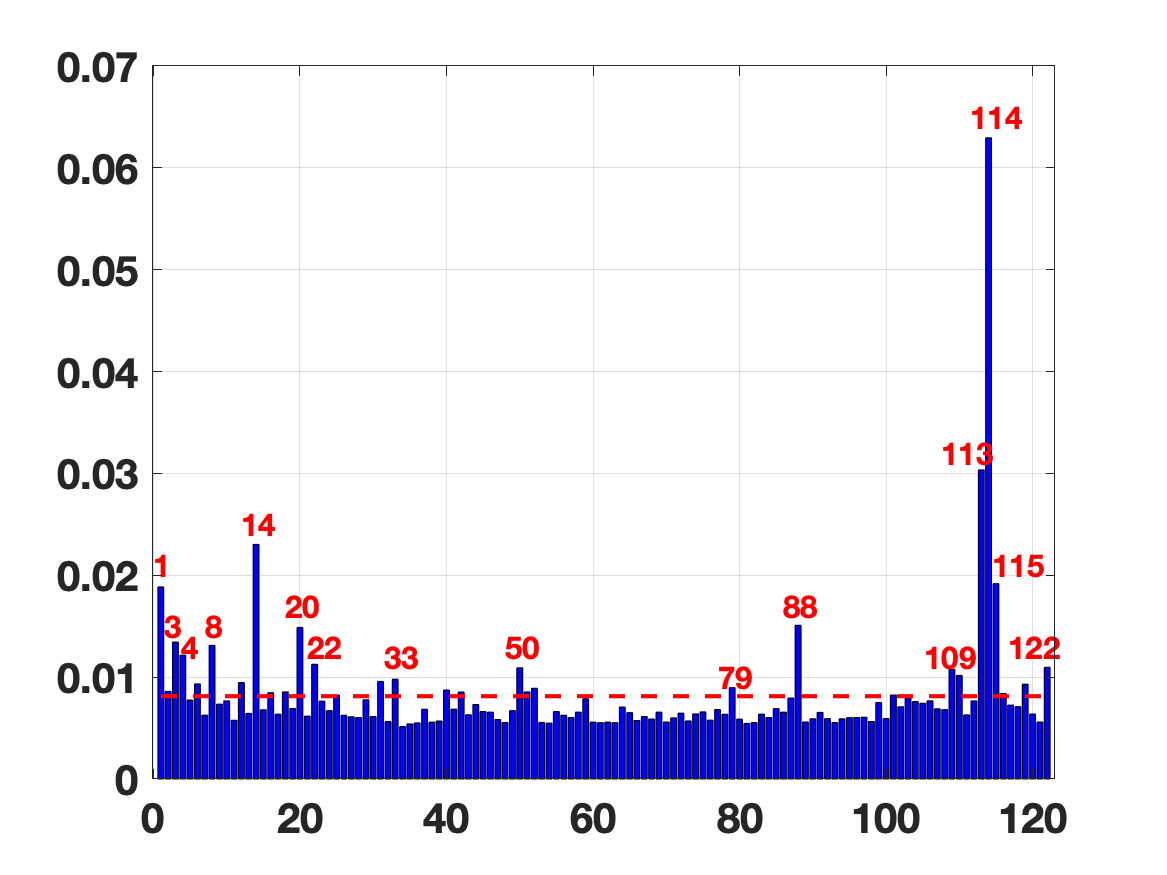}}
		\subfloat[$M=20$]{\includegraphics[width=9cm]{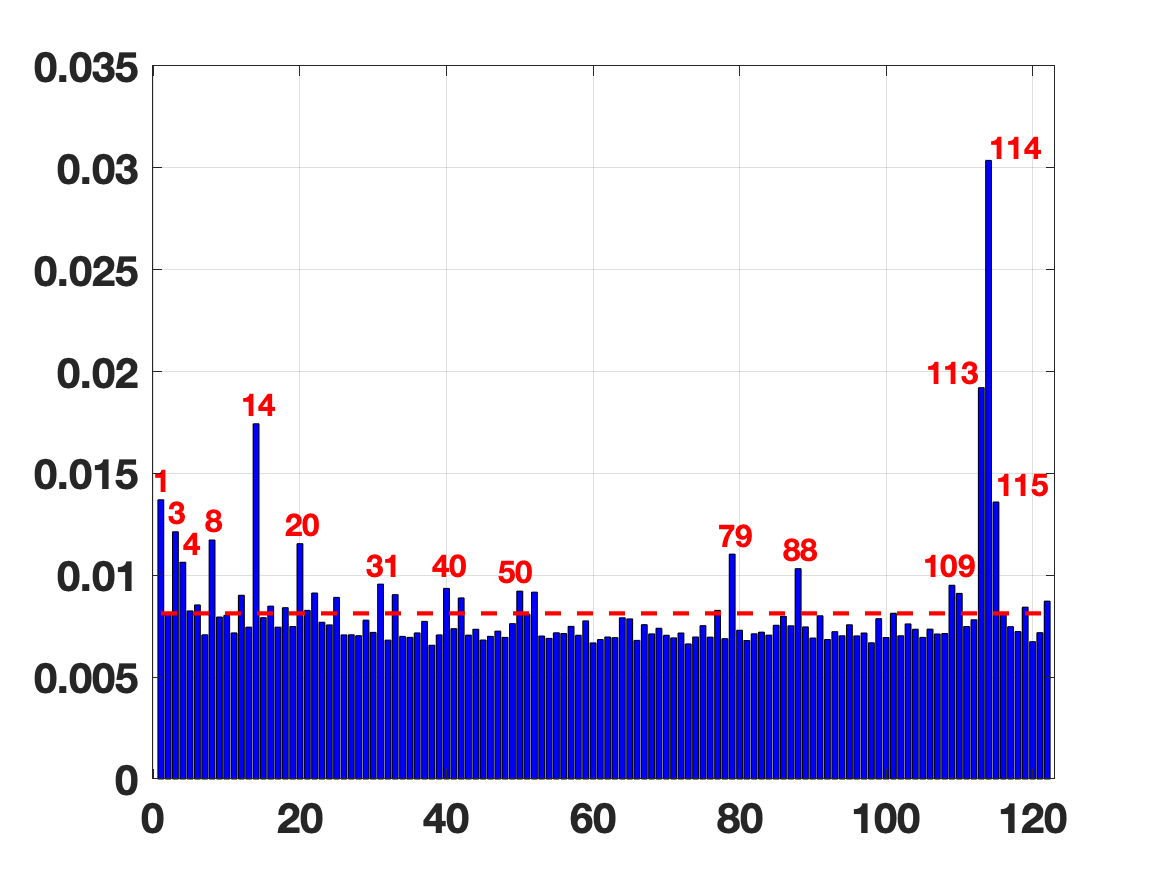}}
		}
	\caption{{\bf (Output 2)} Results in terms of probabilities  obtained by a Gibbs sampling analysis.  The dashed line depicts the uniform discrete distribution with probability $1/122$.}
	\label{fig2Gibbs}
\end{figure*}

\end{document}